\documentclass[a4paper,12pt]{article}
\usepackage{citeref}

\input cohmacros.sty

\advance\textheight3cm        
\advance\topmargin-2.0cm
\advance\textwidth2.6cm
\advance\evensidemargin-1.6cm
\advance\oddsidemargin-1.6cm

\begin{document}

\begin{center}
{\LARGE \bf Foundations of quantum physics} \\[4mm]

{\LARGE \bf II. The thermal interpretation} \\[10mm]

\centerline{\sl {\large \bf Arnold Neumaier}}

\vspace{0.5cm}

\centerline{\sl Fakult\"at f\"ur Mathematik, Universit\"at Wien}
\centerline{\sl Oskar-Morgenstern-Platz 1, A-1090 Wien, Austria}
\centerline{\sl email: Arnold.Neumaier@univie.ac.at}
\centerline{\sl \url{http://www.mat.univie.ac.at/~neum}}

\end{center}


\hfill April 24, 2019

\vspace{0.5cm}


\bigskip
\bfi{Abstract.}
This paper presents the thermal interpretation of quantum physics. 
The insight from Part I of this series that Born's rule has its 
limitations -- hence cannot be the foundation of quantum physics --
opens the way for an alternative interpretation: the thermal 
interpretation of quantum physics.
It gives new foundations that connect quantum physics (including
quantum mechanics, statistical mechanics, quantum field theory and
their applications) to experiment.

The thermal interpretation resolves the problems of the foundations of 
quantum physics revealed in the critique from Part I. It improves the 
traditional foundations in several respects:

\pt
The thermal interpretation reflects the actual practice of
quantum physics, especially regarding its macroscopic implications.

\pt
The thermal interpretation gives a fair account of the interpretational
differences between quantum mechanics and quantum field theory.

\pt
The thermal interpretation gives a natural, realistic meaning to the 
standard formalism of quantum mechanics and quantum field theory in a 
single world, without introducing additional hidden variables.

\pt
The thermal interpretation is independent of the measurement problem.
The latter becomes a precise problem in statistical mechanics rather
than a fuzzy and problematic notion in the foundations.
Details will be discussed in Part III.

\vspace{1cm}

For the discussion of questions related to this paper, please use
the discussion forum \\
\url{https://www.physicsoverflow.org}.

\vspace{2cm}
\tableofcontents 

\vspace{2cm}

\section{Introduction}

\nopagebreak
\hfill\parbox[t]{10.8cm}{\footnotesize

{\em In a statistical description of nature only expectation values
or correlations are observable.
}

\hfill Christof Wetterich, 1997 \cite{Wet}
}

\bigskip

\nopagebreak
\hfill\parbox[t]{10.8cm}{\footnotesize

{\em One is almost tempted to assert that the usual interpretation in 
terms of sharp eigenvalues is 'wrong', because it cannot be 
consistently maintained, while the interpretation in terms of 
expectation values is 'right', because it can be consistently 
maintained.
}

\hfill John Klauder, 1997 \cite[p.6]{Kla}
}

\bigskip

This paper presents the thermal interpretation of
quantum physics. The insight from Part I \cite{Neu.Ifound} of this
series was that Born's rule has its limitations and hence cannot be the
foundation of quantum physics. Indeed, a foundation that starts with
idealized concepts of limited validity is not a safe ground for
interpreting reality.

The analysis of Part I opens the way for an alternative interpretation
 -- the thermal interpretation of quantum physics.
It gives new foundations that connect all of quantum physics (including
quantum mechanics, statistical mechanics, quantum field theory and
their applications) to experiment.

Quantum physics, as it is used in practice, does much more than
predicting probabilities for the possible results of microscopic
experiments. This introductory textbook scope is only the tip of
an iceberg. Quantum physics is used to determine the behavior of
materials made of specific molecules under changes of pressure or
 temperature, their response to external electromagnetic fields
(e.g., their color), the production of energy from nuclear reactions,
the behavior of transistors in the microchips on which modern computers
run, and a lot more. Indeed, it appears to predict the whole of
macroscopic, phenomenological equilibrium and nonequilibrium
thermodynamics in a quantitatively correct way.

Motivated by this fact, this paper defines and discusses a new
interpretation of quantum physics, called the \bfi{thermal 
interpretation}. It is based on the lack of a definite boundary between
the macroscopic and the microscopic regime, and an application of
\sca{Ockham}'s razor \cite{Ock, HofMC},
{\it frustra fit per plura quod potest fieri per pauciora} -- that we
should opt for the most economic model explaining a regularity.

Essential use is made of the fact that everything physicists
measure is measured in a thermal environment for which statistical
thermodynamics is relevant. This is reflected in the characterizing
adjective 'thermal' for the interpretation. The thermal interpretation
agrees with how one interprets measurements in thermodynamics, the
macroscopic part of quantum physics, derived via statistical
mechanics. Extrapolating from the macroscopic case, the thermal
interpretation considers the functions of the state (or of the
parameters characterizing a state from a particular family of states)
as the \bfi{beables}, the conceptual equivalent of objective properties
of what really exists. Some of these are accessible to experiment -- 
namely the expectation values of quantities that have a small 
uncertainty and vary sufficiently slowly in time and space.
Because of the law of large numbers, all thermodynamic variables are in
this category. By its very construction, the thermal interpretation
naturally matches the classical properties of our quantum world.

Section\ref{s.therm} gives a detailed motivation of the thermal 
interpretation and a precise definiton of its basic credo. We introduce 
the Ehrenfest picture of quantum mechanics, the abstract mathematical 
framework used throughout. It describes a closed, deterministic 
dynamics for q-expectations (expectation values of Hermitian operators).
We discuss the ontological status of the thermal interpretation, making 
precise the concept of properties of a quantum system, the concept of 
uncertainty, and the notion of an ensemble. Based on this, we give a 
formal definition of the thermal interpretation, 

In Section \ref{s.thermProb} we consider the way the thermal 
interpretation represents statistical and probabilistic aspects of 
quantum theory. We begin with a discussion of two formal notions of 
classical probability, their relation to the probability concept used 
in applied statistics, and their dependence on the description used. 
We then show how the statistical aspects of the quantum formalism 
naturally follow from the weak law of large numbers. 

In Section \ref{s.qft}, we show that the fact that in relativistic 
quantum field theory, position is a classical parameter while in 
quantum mechanics it is an uncertain quantity strongly affects the 
relation between quantum field theory and reality. Among the beables of 
quantum field theory are smeared field expectations and pair 
correlation functions, which encode most of what is of experimental 
relevance in quantum field theory. We discuss notions of causality and
nonlocality and their relation to the thermal interpretation. 

We also discuss relativistic quantum field theory at finite times, a 
usually much neglected topic essential for a realistic interpretation 
of the universe in terms of quantum field theory. According to the 
thermal interpretation, quantum physics is the basic framework for the 
description of objective reality (including everything reproducible 
studied in experimental physics), from the smallest to the largest 
scales, including the universe as a whole (cf. Subsection 
\ref{ss.universe}). Classical descriptions are just regarded as limiting
cases where Planck's constant $\hbar$ can be set to zero without 
significant loss of quality of the resulting models. 

Except for a brief discussion of the measurement of probabilities in 
Subsection \ref{ss.probMeas}, everything related to the thermal 
interpretation of measurement is postponed to Part III 
\cite{Neu.IIIfound} of this series of papers. There it is shown that 
the thermal interpretation satisfactorily resolves the main stumbling 
blocks in a clear description of the relation between the quantum 
formalism and experimental reality. 

\bigskip

Hints at a possible thermal interpretation of quantum physics go back
at least to 1997; see the above quotes by Wetterich and Klauder.
A recent view closely related to the thermal interpretation is the 
2017 work by \sca{Allahverdyan} et al. \cite{AllBN2}.
The thermal interpretation of quantum physics itself emerged from my
foundational 2003 paper \sca{Neumaier} \cite{Neu.ens}. It was developed 
by me in discussions on the newsgroups de.sci.physik, starting in Spring
2004; for the beginnings see \sca{Neumaier} \cite{Neu.thermBegin}.
A first version of it was fully formalized (without naming the
interpretation) in Sections 5.1, 5.4 and Chapter 7 of the 2008 edition
of the online book by \sca{Neumaier \& Westra} \cite{NeuW}; see also
Sections 8.1, 8.4 and Chapter 10 of the 2011 edition. The name
'thermal interpretation' appeared first in a 2010 lecture
(\sca{Neumaier} \cite{Neu.optslides}).
Later I created a dedicated website on the topic (\sca{Neumaier}
\cite{Neu.thermInt}).

\bigskip

The bulk of this paper is intended to be nontechnical and understandable
for a wide audience being familiar with some traditional quantum
mechanics. For the understanding of the main issues, the knowledge 
of some basic terms from functional analysis is assumed; these are 
precisely defined in many mathematics books. However,quite  a number of 
remarks are addressed to experts and then refer to technical aspects 
explained in the references given.

In the bibliography, the number(s) after each reference give the page 
number(s) where it is cited.

\bigskip
{\bf Acknowledgments.}
Earlier versions of this paper benefitted from discussions with
Hendrik van Hees, Rahel Kn\"opfel and Mike Mowbray.

\section{The thermal interpretation of quantum mechanics}\label{s.therm}

In this section we give a precise definiton of the basic credo of the 
thermal interpretation. Subsection \ref{ss.Ehren} introduces the 
abstract mathematical framework in terms of a Lie product structure
on the set of all q-expectations. This induces a Lie--Poisson bracket 
in terms of which q-expectations behave dynamically as a
classical Hamiltonian system, with a dynamics given by the Ehrenfest 
theorem. This Ehrenfest picture of quantum mechanics is used throughout.
Subsection \ref{ss.properties} discusses the ontological status of the 
thermal interpretation, making precise the concept of properties of a
quantum system. The next two subsections clarify the concepts of 
uncertainty and the notion of an ensemble. The final Subsection 
\ref{ss.formal} then gives a formal definition of the thermal 
interpretation,

\subsection{The Ehrenfest picture of quantum mechanics}\label{ss.Ehren}

As first observed in 1925 by \sca{Dirac} \cite{Dirac1925}, classical 
mechanics and quantum mechanics look very similar when written in 
terms of the Poisson bracket.

Quantities are represented in classical mechanics by functions from a 
space of suitable smooth phase space functions $A(p,q)$, and in quantum 
mechanics by linear operators $A$ on a suitable Euclidean space. 
We define the \bfi{classical Lie product} 
\lbeq{e.cBLie}
A \lp B :=\{B,A\} =\partial_p A\partial_q B-\partial_p B\partial_q A
\eeq
(read $\lp$ as 'Lie')
of classical quantities $A,B$, and the \bfi{quantum Lie bracket}
\lbeq{e.QLie}
A\lp B:= \frac{i}{\hbar}[A,B]=\frac{i}{\hbar}(AB-BA)
\eeq
of quantum mechanical quantities $A,B$. This infix notation is much more
comfortable than the customary bracket notation. In both cases, it is 
easy to verify anticommutativity,
\[
   A\lp B = - B \lp A,
\]
the product rule
\[
   A \lp BC = (A \lp B)C + B(A \lp C),
\]
and the Jacobi relation
\[
   A \lp (B\lp C) = (A \lp B)\lp C + B\lp (A \lp C).
\]
This shows that $\lp$ tuns the space of quantities into a Lie algebra. 
It also shows that the application of $a\lp$ to a quantity behaves 
like differentiation. 

We also write $\int$ both for the Liouville integral 
\lbeq{e.cInt}
\sint A:=\int A(p,q) dp\ dq
\eeq
of a classical quantity $A$ and for the trace 
\lbeq{e.qInt}
\sint A:=\Tr A
\eeq
of a quantum mechanical quantity $A$. With this notation, it is easy to 
verify the invariance under infinitesimal canonical transformations,
\[
\sint A \lp B = 0,
\]
from which one finds the integration by parts formula
\[
\sint (A \lp B)C = \sint A(B \lp C).
\]
In the general theory, for which we refer to \sca{Neumaier} 
\cite{Neu.ens, Neu.qft}, these rules are part of a system of axioms for 
\bfi{Euclidean Poisson algebras}, which allows one to develop 
everything without reference to either the classical or the quantum 
case.

Quantities and linear functionals are, in general, time-dependent;
so we write $\<f\>_t$ for the q-expectation of $f(t)$ at time $t$.
In maximal generality, a \bfi{q-expectation} is written in the form
\lbeq{e.exp}
  \<A\>_t := \sint \rho(t) A(t),
\eeq
where $A(t)$ is an arbitrary time-dependent quantity and $\rho(t)$ a 
time-dependent density operator, a nonnegative Hermitian operator 
normalized by
\[
   \sint \rho = 1.
\]
Their dynamics is given by
\lbeq{esc2}
   \dot A (t) = H_1(t) \lp A(t)
\eeq
for quantities $A$,\footnote{
More generally, if $z$ is a vector of quantities satisfying \gzit{esc2},
quantities given by expressions $A(t)=A(z(t),t)$ with an explicit time
dependence satisfy instead of \gzit{esc2} a differential equation of
the form
\[
   \dot A(t) = i[H_1(t),A(t)] + \partial_t A(z(t),t).
\]
This follows easily from \gzit{esc2} and the chain rule.
The generality gained is only apparent since the numbers
$\<A(z(t),t)\>_t$ are expressible in terms of canonical ones:
 In terms of a Fourier expansion
\[
    A(z(t),t)=\int d \omega e^{i \omega t} A_\omega(z(t)),
\]
we see that
\[
    \<A(z(t),t)\>_t=\int d \omega e^{i \omega t} \<A_\omega(z(t))\>_t,
\]
and the $A_\omega(z(t))$ are canonical quantities respecting
\gzit{esc2}. This allows us to limit the main text to the case
where $A$ has no explicit $t$-dependence.
} 
\lbeq{esc3}
   \dot\rho(t) = \rho(t) \lp H_2(t)
\eeq
for the density operator $\rho$; note the different treatment of 
quantities and the density operator! 
Here $H_1(t)$ and $H_2(t)$ are arbitrary time-dependent
expressions without independent physical meaning; they need
not satisfy the differential equations \gzit{esc2} or \gzit{esc3}.
Integrating \gzit{esc3} shows that $\sint \rho$ is time independent,
so that the dynamics is consistent with the normalization of $\rho$.

\bigskip

As a consequence of the dynamical assumptions \gzit{esc2}-- \gzit{esc3},
the q-expectations \gzit{e.exp} have a deterministic dynamics, given by 
\lbeq{e.Ehrenfest}
\frac{d}{dt} \<A\>_t=\<H\lp A\>_t,
\eeq
We call \gzit{e.Ehrenfest} the \bfi{Ehrenfest equation} since
the special case of this equation where $A$ is a position or momentum 
variable and $H=\frac{p^2}{2m}+V(q)$ is the sum of kinetic and 
potential energy was found in 1927 by \sca{Ehrenfest} \cite{Ehr}. 
Due to the canonical commutation rules, we have 
\lbeq{e.EhrenfestOrig}
\frac{d}{dt} \<q\>_t=\<H\lp q\>_t = \frac{\<p\>_t}{m},~~~
\frac{d}{dt} \<p\>_t=\<H\lp p\>_t = \<-\nabla V(q)\>_t.
\eeq
Note that the Ehrenfest equation does not involve notions of reality or
measurement, hence belongs to the formal core of quantum mechanics
and is valid independent of interpretation issues. 
 
The product rule implies that $\D\frac{d}{dt}\<A\>_t$ only depends on
the sum $H=H_1+H_2$, not on $H_1$ and $H_2$ separately. Thus there is
a kind of gauge freedom in specifying the dynamics, which can be fixed
by choosing either $H_1$ or $H_2$ arbitrarily. Fixing $H_1=0$ (so that
$H_2=H$) makes all quantities $A$ time-independent and defines the
\bfi{Schr\"odinger picture}. Fixing $H_1$ as a reference Hamiltonian
without interactions (so that $H_2=V:=H-H_1$ is the \bfi{interaction})
defines the \bfi{interaction picture}. Fixing $H_2=0$ (so that $H_1=H$)
makes the density operator $\rho$ time-independent and defines the 
\bfi{Heisenberg picture}. In the Heisenberg picture, one finds that 
\lbeq{e.timeShift}
\<\phi(u)\>_s=\<\phi(u+s-t)\>_t
\eeq
for arbitrary times $s,t,u,w,\ldots$.

The Schr\"odinger picture is fully compatible with the formal core of
quantum physics, comprising the postulates (A1)--(A6) discussed in 
Subsection 2.1 of Part I \cite{Neu.Ifound}. In particular, the von 
Neumann equation
\lbeq{e.vN}
\frac{d}{dt}\rho(t) = \frac{i}{\hbar} [\rho(t),H] \for t \in [t_1,t_2],
\eeq
holds for closed systems, giving a deterministic dynamics for the
density operator. As discussed in Subsection 2.3 of Part I
\cite{Neu.Ifound}, one obtains the Schr\"odinger equation in the
limit of pure states.

In place of the traditional Heisenberg, Schr\"odinger, and interaction 
pictures, one can also consider another equivalent picture, in which 
only q-expectations figure as dynamical variables. 
The name \bfi{Ehrenfest picture}
is suggestive since, for the standard multiparticle Hamiltonian
and $f=p,q$, this reduces to the Ehrenfest equation.
In terms of the Lie bracket on q-expectations defined by the formula
\lbeq{e.Lie}
\<A\> \lp \<B\> := \<A\lp B\>,
\eeq
the family of q-expectations becomes a Lie algebra $\Lz$, and
the Ehrenfest equation \gzit{e.Ehrenfest} becomes
\lbeq{e.Edyn}
\frac{d}{dt} \<A\> = \<H\> \lp \<A\>.
\eeq
Equation \gzit{e.Edyn} is quite remarkable as it is manifestly 
independent of how $H$ is split and how the time-dependent expectation 
is expressed as \gzit{e.exp}.

The Ehrenfest picture gives a complete picture of the 
(classical or) quantum kinematics and a deterministic dynamics for the 
q-expectations that is equivalent to the Schr\"odinger picture, the 
Heisenberg picture, and the interaction pictures.

It is interesting to interpret the above in terms of Hamiltonian 
dynamics on Poisson manifolds. An extensive discussion of classical 
Hamiltonian dynamics on Poisson manifolds, in particular using 
Lie--Poisson brackets, and its application to rigid rotors and fluid 
dynamical systems is given in  \sca{Marsden \& Ratiu} \cite{MarR}.
A \bfi{Poisson manifold} is a smooth manifold together with a Lie 
product on $\Ez=C^\infty(M)$ that turns $\Ez$ into a commutative 
Poisson algebra. Associated with the Lie algebra $\Lz$ of 
q-expectations is the manifold $\Lz^*$ of continuous linear functionals 
on $\Lz$. On $\Ez=C^\infty(\Lz^*)$, a Lie product is given by the 
classical Lie--Poisson bracket which canonically extends the formula 
\gzit{e.Lie} to smooth functions of q-expectations. This turns 
$\Ez$ into a commutative Poisson algebra, hence $\Lz^*$ into a Poisson 
manifold.\footnote{
In terms of the symplectic Poisson algebra considered by \sca{Strocchi} 
\cite{Stro} to express the quantum mechanical dynamics of pure states 
$\psi$, the present Poisson algebra corresponds to the Poisson
subalgebra of even functions of $\psi$.
} 
In these terms, the Ehrenfest picture of {\em quantum} mechanics is 
just {\em classical} (but nonsymplectic) Hamiltonian dynamics in the 
Poisson manifold $\Lz^*$, with expected energy $\<H\>$ as the 
{\em classical} Hamiltonian. In particular, as can also be seen 
directly, expected energy is conserved.

\subsection{Properties}\label{ss.properties}

The \bfi{beables} of a physical system in the sense of \sca{Bell}
\cite[p.41]{Bell} are the linear functionals of its density operator 
$\rho$, called the \bfi{state} of the system, and any complex-valued 
function constructed from these. The linear functionals are in 1-to-1 
correspondence with \bfi{q-expectations}, i.e., expressions of the 
form\footnote{
Note that the density operator, viewed as a time-dependent object, is
picture-dependent, but with the corresponding time-dependence of the 
linear operator $A$ as discussed in Subsection \ref{ss.Ehren},
the q-expectations are picture-independent.
} 
\lbeq{e.uncVal}
\ol A=\<A\>:=\Tr\rho A,
\eeq
where $A$ is a linear operator (or more generally a sesquilinear form)
on the Euclidean space $\Hz$. By definition, all functions of 
q-expectations are beables, and all beables arise in this way. T
his gives a clear formal meaning to the 
notion of existence, an \bfi{ontology}: In the thermal interpretation,
something is said to \bfi{exist}, to be \bfi{real}, and to be 
\bfi{objective} -- three ways of expressing the same --, if and only if 
it can be expressed in terms of beables only.

As observed in 1927 by \sca{von Neumann} \cite[p.255]{vNeu1927b} -- 
who was the first to base quantum mechanics upon expectations rather 
than probabiilties --, the specification of all expectations determines 
the complete state (density operator) $\rho$. Thus every function of 
the state $\rho$ can be rewritten (in any fixed picture) as a function 
of q-expectations, hence is a beable. 

Only a small set of beables are practically (approximately) observable.
Clearly, anything computable in quantum physics belongs there.
Whenever we are able to compute something from raw measurements
according to the rules of some meaningful protocol, and it adequately 
agrees with something derivable from quantum physics, we call the 
result of that computation a \bfi{measurement} of the latter. 
This correctly describes the practice of measurement in its most 
general form. Formal details will be given in Part III 
\cite{Neu.IIIfound}.

We recall the rules (S1)--(S3) from Subsection 3.4 and (R) from 
Subsection 4.1 of Part I \cite{Neu.Ifound} that we 
found necessary for a good interpretation:

{\bf (S1)}
The state of a system (at a given time) encodes everything that can be
said about the system, and nothing else.

{\bf (S2)}
Every property of a subsystem is also a property of the whole system.

{\bf (S3)} The state of a system determines the state of all its
subsystems.

{\bf (R)} 
Something in real life 'is' an instance of the theoretical concept
if it matches the theoretical description sufficiently well.

The problems that traditional interpretations have with the relation
between the state of a system and that of a subsystem were discussed
in Subsection 3.4 of Part I \cite{Neu.Ifound}. They are resolved
in the thermal interpretation. Indeed, the use of density operators as
states implies that the complete state of a system completely and
deterministically specifies the complete state of every subsystem.

Rule (S1) holds in the thermal interpretation because everything that
can be computed from the state is a beable.

The other rules are satisfied in the thermal interpretation by making
precise the notions of 'statement about a system' and
'property of a system'.

A \bfi{statement} is a $\{{\tt true},{\tt false}\}$-valued function of
the state.
A \bfi{property} of a system at time $t$ is a statement $P$ such that
$P(\rho(t))$ is true, where $\rho(t)$ is the state of the system at
time $t$.

A \bfi{subsystem} of a system is specified by a choice declaring some
of the quantities (q-observables) of the system to be the distinguished
quantities of the subsystem. This includes a choice for the
Hamiltonian of the subsystem. The dynamics of the subsystem is generally
not closed, hence not given by the Ehrenfest equation \gzit{e.Edyn}.
However, in many cases, an approximate closed dynamical description is 
possible; this will be discussed in more detail in Part III 
\cite{Neu.IIIfound}.

Note that unlike in traditional interpretations, no tensor product
structure is assumed. However, suppose that the latter is present,
$\Hz=\Hz_S\otimes \Hz_\env$, and the quantities of the subsystem are the
linear operators of $\Hz_S\tensor 1$. Then, without changing any of the
predictions for the subsystem, the Hilbert space of the subsystem may be
taken to be the smaller Hilbert space $\Hz_S$, and the quantities of
the subsystem are the linear operators of  $\Hz_S$. Then the density
operator of the subsystem is the \bfi{reduced state} obtained as the
partial trace over the environment Hilbert space $\Hz_\env$.
In this sense, (S3) holds.

Rule (R) just amounts to a definition of what it means of something in
real life to 'be an $X$', where $X$ is defined as a theoretical
concept.

\subsection{Uncertainty}

\nopagebreak
\hfill\parbox[t]{10.8cm}{\footnotesize

{\em A quantity in the general sense is a property ascribed to
phenomena, bodies, or substances that can be quantified for, or
assigned to, a particular phenomenon, body, or substance. [...]
The value of a physical quantity is the quantitative expression
of a particular physical quantity as the product of a number and a
unit, the number being its numerical value.
}

\hfill Guide for the Use of the International System of Units
(\sca{Taylor} \cite{Tay.SI})
}

\bigskip

\nopagebreak
\hfill\parbox[t]{10.8cm}{\footnotesize

{\em The uncertainty in the result of a measurement generally consists 
of several components which may be grouped into two categories according
to the way in which their numerical value is estimated.
\\
\phantom{\spc} Type A. Those which are evaluated by statistical methods
\\
\phantom{\spc} Type B. Those which are evaluated by other means
\\{}[...]
The quantities $u_j^2$ may be treated like variances and the quantities 
$u_j$ like standard deviations. 
}

\hfill NIST Reference on Constants, Units, and Uncertainty \cite{SI2}
}

\bigskip

Uncertainty permeates
all of human culture, not only science. Everything quantified by real 
numbers (as opposed to counting) is intrinsically uncertain because we
cannot determine a real number with arbitrary accuracy. Even counting 
objects or events is uncertain in as much the criteria that determine 
the conditions under which something is counted are ambiguous. 
(When does the number of people in a room change by one while someone 
enters the door?) 

The thermal interpretation of quantum physics takes the virtually 
universal presence of uncertainty as the most basic fact of science and 
gives it a quantitative expression. Some of this uncertainty can be 
captured by probabilities and statistics, but the nature of much of 
this uncertainty is conceptual. Thus uncertainty is a far more basic 
phenomenon than statistics. It is an uncertainty in the notion of 
measurability itself. What does it mean to have measured something? 

To be able to answer this we first need clarity in the terminology.
To eliminate any trace of observer issues\footnote{
Except when relating to tradition, we deliberately avoid the notion of
observables, since it is not clear on a fundamental level what it
means to `observe' something, and since many things (such as the fine
structure constant, neutrino masses, decay rates, scattering cross
sections) observable in nature are only indirectly related to what is
traditionally called an `observable' in quantum physics.
} 
from the terminology, we use the word \bfi{quantity} (as recommended
in the above quote from the Guide to the International System of Units)
or -- in a more technical context -- \bfi{q-observable}\footnote{
\label{f.q}
Note that renaming notions has no observable consequences, but strongly
affects the interpretation. To avoid confusion, I follow here as in
Part I \cite{Neu.Ifound} the convention of \sca{Allahverdyan} et al. 
\cite{AllBN2} and add the prefix 'q-' to all traditional quantum 
notions that get here a new interpretation and hence a new terminology. 
In particular, we use the terms q-observable, q-expectation, q-variance,
q-standard deviation, q-probability, q-ensemble for the conventional 
terms observable, expectation, variance, standard deviation, 
probability, and ensemble.
} 
whenever quantum tradition uses the word observable. 
Similarly, to eliminate any trace of a priori statistics from the
terminology, we frequently use the terminology \bfi{uncertain value} (in
\cite{NeuW} simply called value) instead of q-expectation value,
and \bfi{uncertainty} instead of q-standard deviation.

For the sake of definiteness consider the notion of uncertain position.
This may mean two things.

1. It may mean that the position could be certain, as in classical
Newtonian physics, except that we do not know the precise value.
However, measurements of arbitrary accuracy are at least conceivable.

2. It may mean that the position belongs to an extended object, such as
a neutron star, the sun, a city, a house, a tire, or a water wavelet.
In this case, there is a clear approximate notion of position, but it
does not make sense to specify this position to within millimeter
accuracy.

It seems to be impossible to interpret the second case in terms of the
first in a natural way. The only physically distinguished point-like
position of an extended object is its
center of mass. Classically, one could therefore think of defining the
exact position of an extended object to be the position of its center
of mass. But the sun, a city, a house, or a water wavelet do not even
have a well-defined boundary, so even the definition of their center
of mass, which depends on what precisely belongs to the object, is
ambiguous. And is a tire really located at its center of mass -- which
is well outside the material the tire is made of? Things get worse in
the microscopic realm, where the center of mass of a system of quantum
particles has not even an exactly numerically definable meaning.

On closer inspection it seems that the situation of case 2 is very
frequent in practice. Indeed, it is the {\em typical} situation in the
macroscopic, classical world. Case 1 appears to be simply a convenient
but unrealistic idealization.

The uncertainty in the position of macroscopic objects such
as the sun, a city, a house, a tire, or a water wavelet is therefore a
conceptual uncertainty impossible to resolve by measurement. The
thermal interpretation asserts that quantum uncertainty is an
uncertainty of the same conceptual kind.

Thus uncertainty is only partially captured through statistical
techniques. The latter apply only in case of highly repetitive 
uncertain situations, leading to a particular kind of uncertainty 
called aleatoric uncertainty (see, e.g., \cite{DerD,Pat}). More general 
kinds of uncertainty are discussed in the NIST Reference on Constants, 
Units, and Uncertainty \cite{SI2}, which may be regarded as the de facto
scientific standard for representing uncertainty. This source explicitly
distinguishes between uncertainties ''which are evaluated by statistical
 methods'' and those ''which are evaluated by other means''. For the
second category, it is recognized that the uncertainties are not
statistical but should be treated ''like standard deviations''.

The thermal interpretation follows this pattern by explicitly
recognizing that not all uncertainty can be expressed statistically,
though it is expressed in terms of uncertainty formulas that behave
like the corresponding statistical concepts.

\subsection{What is an ensemble?}\label{ss.ensemble}

\nopagebreak
\hfill\parbox[t]{10.8cm}{\footnotesize

{\em We may imagine a great number of systems of the same nature, but
differing in the configurations and velocities which they have at a
given instant, and differing not merely infinitesimally, but it may be
so as to embrace every conceivable combination of configuration
and velocities. [...] The first inquiries in this field were indeed
somewhat narrower in their scope than that which has been mentioned,
being applied to the particles of a system, rather than to independent
systems.
}

\hfill Josiah Willard Gibbs, 1902 \cite[pp. vii--viii]{Gib}
}

\bigskip

\nopagebreak
\hfill\parbox[t]{10.8cm}{\footnotesize

{\em So aufgefa{\ss}t, scheint die Gibbssche Definition geradezu 
widersinnig. Wie soll eine dem K\"orper wirklich eignende Gr\"o{\ss}e 
abhangen nicht von dem Zustand, den er hat, sondern den er 
m\"oglicherweise haben k\"onnte? [...] Es wird eine Gesamtheit 
mathematisch fingiert [...] erscheint es schwierig, wenn nicht 
ausgeschlossen, dem Begriffe der kanonischen Gesamtheit eine 
physikalische Bedeutung abzugewinnen.
}

\hfill Paul Hertz, 1910 \cite[p.226f]{Her} 
}

\bigskip

The thermal interpretation of quantum physics says that, consistent
with statistical thermodynanics, a q-expectation (q-ensemble mean)
is interpreted as an in principle approximately measurable quantity.
Except when the statistical context is immediate (such as in computer
simulations), the q-expectation should not be interpreted as a
statistical average over a population\footnote{
Physicists usually speak of an ensemble in place of a population.
In this paper, the statistical term \bfi{population} is used instead, 
to keep the discussion unambiguous, since in connection with the 
microcanonical, canonical, or grand canonical ensemble the term 
\bfi{ensemble} is essentially synonymous with a density operator of a 
particularly simple form.
} 
of many realizations.

The q-expectation, conventionally called the ensemble expectation,
becomes in the thermal interpretation simply the uncertain value.

Therefore, the notion of q-ensemble is to be understood not as an actual
repetition by repeated preparation. It should be understood instead
in the sense of a fictitious collection of imagined copies of which
{\em only one} is actually realized --, giving an intuitive excuse for
using the statistical formalism for a single system.

The association of a ficticious ensemble to {\em single} thermal systems
goes back to Gibbs, the founder of the ensemble approach to classical
statistical mechanics. He was very aware that thermodynamics and hence
statistical mechanics applies to single physical systems.
His arguments are today as cogent as when he introduced them.

In classical statistical mechanics, the distinction between the
deterministic and stochastic description becomes blurred, as {\it each
single} macroscopic system is already described by a phase space
density (multiparticle distribution function), although the latter
behaves mathematically in every respect like a probability density that
expresses the properties of a population of identical systems.

This tension in the terminology is already visible in the famous
statistical mechanics textbook by \sca{Gibbs} \cite{Gib} in 1902, where
he introduced in the preface (from which the above quote is taken) 
fictitious ensembles to bridge the conceptual gap.

Thus to deduce properties of macroscopic materials, Gibbs uses an 
ensemble of macroscopic systems -- in contrast to Boltzmann, who 
introduced statistical mechanics for gases by using ensembles of 
microscopic atoms. 
Treating a collection of particles in a gas as an ensemble (as
Boltzmann did) makes statistical sense as there are a huge number of
them. But the Gibbs formalism is applied to single
macroscopic systems such as a brick of iron rather than to its many
constituents. Treating a single system as part of a fictitious ensemble
was a very bold step taken by Gibbs \cite[p.5]{Gib}:
{\it ''Let us imagine a great number of independent systems, identical
in nature, but differing in phase, that is, in their condition with
respect to configuration and velocity.''} 
This allowed him to extend Boltzmann's work from ideal gases to
arbitrary chemical systems, in a very robust way. His statistical
mechanics formalism, as encoded in the textbook
(\sca{Gibbs} \cite{Gib}), survived the quantum revolution almost without
change -- the book reads almost like a modern book on statistical
mechanics!

Though exceeedingly successful, Gibbs' fictitious ensemble raised in
his time severe objections in the physics community, such as the 
response by Hertz quoted above, who complained that an ensemble is 
feigned mathematically. Of course Gibbs was aware that imagined systems 
have no physical implications, but these were needed at a time where
mathematics had not yet the abstract character that it has today.

Today, mathematical theories are simply formal systems used without
hesitation in applications in which the terms may mean something
completely different from their meaning in the uses that gave the names
to the terms.
For example, the mathematical notion of a vector is today an abstract
tool routinely used in contexts where the original geometric notion of
a vector is meaningless: No physicist thinks of a quantum mechanical
state vector in terms of a little arrow depicting a translational
motion.

In the same spirit, mathematical statistics (and hence statistical
mechanics) may be used as a tool in which expectation values figure as
abstract notions without the need to imagine an ensemble of copies of
the single system under study over which the expectation would be an
imagined average. Thus we are liberated from having to think of the
mathematical q-expectation values manipulated in statistical mechanics
as being true averages over fictitious copies without a physical
meaning.
Instead, the statistical terminology is simply a reminder of which laws
(originally stemming from statistical data analysis) are applicable to
these values, in the same way as the geometric terminology of a vector
indicates the laws valid for manipulating objects behaving
algebraically like vectors.

\subsection{Formal definition of the thermal interpretation}
\label{ss.formal}

\nopagebreak
\hfill\parbox[t]{10.8cm}{\footnotesize

{\em Wenn wir aus jenem mathematischen Schema physikalische Resulate
ableiten wollen, so m\"ussen wir den quantentheoretischen Gr\"o{\ss}en,
 also den Matrizen (oder 'Tensoren' im mehrdimensionalen Raum)
Zahlen zuordnen. [...] Man kann also sagen: Jeder quantentheoretischen
Gr\"o{\ss}e oder Matrix l\"a{\ss}t sich eine Zahl, die ihren 'Wert'
angibt, mit einem bestimmten wahrscheinlichen Fehler zuordnen; der
wahrscheinliche Fehler h\"angt vom Koordinatensystem ab; f\"ur jede
quantentheoretische Gr\"o{\ss}e gibt es je ein Koordinatensystem, in
dem der wahrscheinliche Fehler f\"ur diese Gr\"o{\ss}e verschwindet.
Ein bestimmtes Experiment kann also niemals f\"ur alle
quantentheoretischen Gr\"o{\ss}en genaue Auskunft geben
}

\hfill Werner Heisenberg, 1927 \cite[p.181f]{Hei1927}
}


\bigskip

\nopagebreak
\hfill\parbox[t]{10.8cm}{\footnotesize

{\em Wir hatten ja immer leichthin gesagt: die Bahn des Elektrons in 
der Nebelkammer kann man beobachten. Aber vielleicht war das, was man 
wirklich beobachtet, weniger. Vielleicht konnte man nur eine diskrete 
Folge von ungenau bestimmten Orten des Elektrons wahrnehmen. 
Tats\"achlich sieht man ja nur einzelne Wassertr\"opfchen in der Kammer,
die sicher sehr viel ausgedehnter sind als ein Elektron. 
Die richtige Frage mu{\ss}te also lauten: Kann man in der 
Quantenmechanik eine Situation darstellen, in der sich ein Elektron
ungef\"ahr -- das hei{\ss}t mit einer gewissen Ungenauigkeit -- 
an einem gegebenen Ort befindet und dabei ungef\"ahr -- das hei{\ss}t 
wieder mit einer gewissen Ungenauigkeit -- eine vorgegebene 
Geschwindigkeit besitzt, und kann man diese Ungenauigkeiten so gering 
machen, da{\ss} man nicht in Schwierigkeiten mit dem Experiment 
ger\"at?
}

\hfill Werner Heisenberg, 1972 \cite[p.77f]{Hei.Teil}
}

\bigskip

The thermal interpretation of quantum physics uses for the description 
of quantum physics a formal framework consisting of
\begin{itemize}
\item
a Euclidean space $\Hz$ on which \bfi{states} are encoded by density 
operators, positive semidefinite linear trace 1 operator $\rho$ from 
$\Hz$ to itself (in the most regular case; to its completion in the 
general case);
\item
a representation of the standard model of the electromagnetic, weak, 
and strong interactions plus some form of gravity (not
yet fully known) to describe the fundamental field content;
\item
a unitary representation of the Heisenberg, Galilei or Poincar\'e 
group\footnote{
Except for a few passing remarks concerning gravity, we assume in this 
paper a flat spacetime. In a quantum field theory of gravity, we'd 
also need something like a unitary representation of the diffeomorphism 
group of the spacetime manifold, which has its own foundational 
problems.
} 
to account for conservative dynamics and the principle of relativity in 
its nonrelativistic or special relativistic situation, respectively;
\item
and the Ehrenfest picture for the dynamics of q-expectations.
\end{itemize}

The representation defines the basic physical quantities (linear
operators on $\Hz$) characterizing the system -- energy, momentum,
angular momentum. $\Hz$ itself is a dense subspace of the physical
Hilbert space and forms a common domain for the basic physical
quantities.

As described in Subsection \ref{ss.properties} above, the state defines 
the properties a particular instance of the system has, and hence what 
exists in the system.

The thermal interpretation avoids both the philosophically problematic
notion of probability, and the anthropomorphic notions of knowledge and
measurement. (We shall reconsider the notion of probability in 
Section \ref{s.thermProb}, that of knowledge in Subsection 
\ref{ss.condProb}, and that of measurement in Part III 
\cite{Neu.IIIfound}.) Instead, it gives the q-expectation value
an interpretation in terms of a fundamental -- not further explained but
intuitive -- notion of uncertainty, generalizing statistical practice:

\bfi{(GUP)} \bfi{General uncertainty principle:}
{\it A Hermitian\footnote{
The uncertain value and its uncertainty makes also sense in the 
nonhermitian case, but the uncertainty must be defined in this case as
\[
\sigma_A:=\sqrt{\Big\<(A-\ol A)^*(A-\ol A)\Big\>}
=\sqrt{\<A^*A\>-|\ol A^2|}.
\]
} 
quantity $A$ has the \bfi{uncertain value}
$\ol A=\<A\>$ with an \bfi{uncertainty} of\footnote{
The equivalence of both expressions defining $\sigma_A$ follows from
$\ol A=\<A\>$ and 
\[
\<(A-\ol A)^2\>=\<A^2-A\ol A-\ol AA+\ol A^2\>
=\<A^2\>-\<A\>\ol A-\ol A\<A\>+\ol A^2=\<A^2\>-\ol A^2.
\]
} 
\lbeq{e.sigmaA}
\sigma_A:=\sqrt{\Big\<(A-\ol A)^2\Big\>}=\sqrt{\<A^2\>-\ol A^2}.
\eeq
In particular, the uncertain value $\ol A$ is informative whenever its
uncertainty $\sigma_A$ is much less than $|\ol A|$ } 

This contrasts with the traditional interpretations, which give the
q-expectation value an interpretation in terms of a not further
explained, fundamental notion of random measurement or a not further
explained, fundamental notion of knowledge (or information), with all
their associated foundational difficulties that were explored since 
Born proposed his interpretation. 

Much of quantum physics can be developed without using the notion of 
probability at all. The notion of uncertain values $\<A\>$ suffices for 
almost all of quantum chemistry and quantum statistical mechanics. 
For equilibrium statistical mechanics, this can be seen from the 
treatment in \sca{Neumaier \& Westra} \cite{NeuW}.

\bigskip

As discovered by \sca{Heisenberg} \cite{Hei1927}, quantum physics 
predicts its own uncertainty. The q-variances of corresponding 
components of position $q$ and momentum $p$ cannot be both arbitrarily 
small since
\[
\sigma_{p_j}\sigma_{q_j}\ge \half \hbar.
\]
This inequality is the famous \bfi{Heisenberg uncertainty relation}
-- not to be confused with the general uncertainty principle (GUP) 
defined above. The Heisenberg uncertainty relation is a special case of 
the more general statement (due to \sca{Robertson} \cite{Rob}) that for 
non-commuting Hermitian operators $A,B$,
\lbeq{e.uncRel}
\sigma_A\sigma_B\ge \half |\<[A,B]\>|,
\eeq
which follows from the definitions.\footnote{
Indeed, the relation remains unchanged when subtracting from $A$ and
$B$ its q-expectation, hence it suffices to prove it for the case
where both q-expectations vanish. In this case, $\<A^2\>=\sigma_A^2$ and
$\<B^2\>=\sigma_B^2$, and the Cauchy--Schwarz inequality gives
$|\<AB\>|^2\le \<A^2\>\<B^2\>=\sigma_A^2\sigma_B^2$, hence
$|\<AB\>|\le\sigma_A\sigma_B$. On the other hand, one easily checks that
$i\im\<AB\>=\half\<[A,B]\>$, so that
$\half |\<[A,B]\>|=|\im\<AB\>|\le |\<AB\>|$. Combining both inequalities
gives the assertion.
} 
Like Ehrenfest's equation, the Heisenberg uncertainty relation does not 
involve notions of reality or measurement, hence belongs to the formal 
core of quantum mechanics. Both are valid independent of the 
interpretation of the q-expectations and the q-variances, respectively.

As discussed in Part I \cite{Neu.Ifound} (and exemplified in more detail
in Part III \cite{Neu.IIIfound} and Part IV \cite{Neu.IVfound}), the 
\bfi{interpretation}, i.e., the identification of formal properties 
given by uncertain values with real life properties of a physical 
system, is done by means of the generalized uncertainty principle (GUP),
interpreting the q-expectations as beables and the q-variances as their 
uncertainty, together with

\bfi{(CC)} \bfi{Callen's criterion} (\sca{Callen} \cite[p.15]{Cal}):
{\it Operationally, a system is in a given state if its properties are 
consistently described by the theory for this state.
} 

This is enough to find out in each single case how to approximately 
measure the uncertain value of a quantity of interest, though it may 
require considerable experimental ingenuity to do so with low 
uncertainty. The uncertain value $\ol X$ is considered informative
only when its uncertainty $\sigma_X$ is much less than $|\ol X|$.

This interpretation gives both Ehrenfest's equation and the Heisenberg 
uncertainty relation their practical importance: Ehrenfest's equation 
leads to the emergence of classical mechanics as approximate dynamics in
cases where the variances are tiny (see Part IV \cite{Neu.IVfound} for 
details). Heisenberg's uncertainty relation enforces an unavoidable 
limit on the accuracy with which position and momentum can be specified 
simultaneously. 

In the thermal interpretation, even pointlike quantum objects are 
extended: Every pointlike quantum object has a 3-component position 
vector $\q$, hence is extended to the extent determined by the 
computable position uncertainty 
$\sigma_\q=\sqrt{\sigma_{\q_1}^2+\sigma_{\q_2}^2+\sigma_{\q_3}^2}$,
where $\q$ is the position vector of the object. In spacetime, the
uncertain positions $\<\q\>_t$  traced out an uncertain world line,
and the quantum object can be visualizied as moving along a 
nested family of fuzzy \bfi{world tubes}, the union of the intervals 
$[\<\q\>_t -\kappa\sigma_{\q}(t),\<\q\>_t +\kappa\sigma_{\q}(t)]$,
for reasonable $\kappa$ of order one.

Thus the thermal interpretation answers Heisenberg's question from 
the early days of quantum mechanics quoted above, 
{\it ''Can quantum mechanics represent the fact that an electron finds 
itself approximately in a given place and that it moves approximately 
with a given velocity, and can we make these approximations so close 
that they do not cause experimental difficulties?''}
It gives the electrons real but uncertain paths, making them move along 
an extended world tube rather than an infinitesimally thin world line.

\section{Thermal interpretation of statistics and probability}
\label{s.thermProb}

In this section we consider the way the thermal interpretation 
represents statistical and probabilistic aspects of quantum theory.
We begin in Subsections \ref{ss.probViaEx} and \ref{ss.condProb}
with a discussion of two formal notions of classical probability, their 
relation to the probability concept used in applied statistics, and
their dependence on the description used. 

Historically, the concept of classical probability (including its use in
stochastic processes) was given an undisputed formal mathematical 
foundation in 1933 in terms of the measure-theoretic setting of 
\sca{Kolmogorov} \cite{Kol}. Apart from this traditional axiomatic 
foundation of classical probability theory there 
is a less well-known equivalent axiomatic treatment by \sca{Whittle} 
\cite{Whi} in terms of expectations. Here probabilities appear as the 
expectations of statements, $\{0,1\}$-valued random variables. 
\sca{von Plato} \cite{vPla} discusses the history of the concept of
probability. \sca{Kr\"uger} et al. \cite{KruGM} discuss the history of 
probability in the various fields of application. \sca{Sklar} \cite{Skl}
discusses the philosophical problems of the probability concept, with an
emphasis on statistical mechanics. 

Subsection \ref{ss.stat} then shows how the statistical aspects of the
quantum formalism naturally follow from the weak law of large numbers. 
The notions of c-probability in the classical case and of q-probability 
in the quantum case are formally defined in Subsection 
\ref{ss.whatIsProb}; they are related to experiment in Subsection 
\ref{ss.probMeas}. The final subsection briefly discusses various
examples where probabilistic features emerge from fully determinstic 
situations -- a theme to be taken up in much more detail in Part III
\cite{Neu.IIIfound}.

\subsection{Classical probability via expectation}\label{ss.probViaEx}

This subsection gives an elementary introduction to classical 
probability theory along the lines of \sca{Whittle} \cite{Whi}, 
similar in spirit to the formal core of quantum mechanics.

Let $\Omega$ denote a finite or infinite set of conceivable
\bfi{experiments} $\omega$.\footnote{
In probability theory $\Omega$ is called the \bfi{sample space}.
In statistics, the $\omega\in\Omega$ may be identified with actual 
experiments carried out in the past or in the future. In classical 
statistical mechanics, the $\omega$ are only hypothetical experiments 
from a fictitious ensemble in the sense of Gibbs, as discussed in 
Subsection \ref{ss.ensemble}. In general, $\Omega$ is just a set, and 
calling the  $\omega\in\Omega$ experiments is just a convenient 
intuition without formal meaning (comparable to calling elements of a 
vector space vectors even though they may be functions or matrices).
For examples, see Subsection \ref{ss.stochDet}.
} 
Let $\Ez$ be a vector space of real valued
functions on $\Omega$ containing the constant functions.
We identify constant functions with their function values.
The elements $A\in\Ez$ are called (real) \bfi{random variables}, and the
value $A(\omega)$ is called the \bfi{realization} of $A$ in experiment
$\omega$.
The procedure that defines how to obtain the realization $f(\omega)$ for
any experiment $\omega$ is called the \bfi{protocol} defining the
random variable $f$. Functions, order relations, unary and binary
operations, and limits on random variables are defined pointwise;
note, however, that a pointwise function $f(A)$ of a random variable
$A\in\Ez$ does not necessarily lie in $\Ez$.

A \bfi{sample} is a finite set $S$ of $|S|>0$ experiments; the
associated \bfi{sample mean} of a random variable $A$ is defined by
\lbeq{e.sample}
\< A \>_S:=\frac{1}{|S|}\sum_{\omega\in S} A(\omega).
\eeq
It is easily checked that any sample mean $\<\cdot\>=\<\cdot\>_S$
satisfies the following rules:

(E1)~ $\<1\>=1$.\\
(E2)~ $\<\alpha A+\beta B\>=\alpha\<A\>+\beta\<B\>$ for
      $\alpha,\beta\in\Rz$.\\
(E3)~ $A\ge 0$ implies $\< A\>\geq 0$.\\
(E4)~ $A\ge 0$, $\< A\>=0$ implies $A=0$.\\
(E5)~ $A_{k}\downarrow 0$ implies $\< A_{k} \>\downarrow 0$.

Here $\downarrow$ denotes pointwise convergence from above.
To abstract from a particular sample we define a \bfi{stochastic model}
as an arbitrary mapping that assigns to each random variable $A$ a real
number $\<A\>$, called the \bfi{expectation} (or \bfi{expected value}
or \bfi{mean}) of $A$  such that the axioms (E1)--(E5) hold whenever
the expectations in question exist. All samples, arbitrary convex
combinations of samples, and their limits, define a stochastic model.
Given real statistical data from real experiments, the quality of a
stochastic model is assessed by how well the expectations of key random
variables match corresponding sample expectations for samples drawn at
random in some informal sense.

As a simple consequence of the axioms, we note:

(E6)~$\<A^{2}\>= 0$ implies $A=0$,\\
(E7)~$A\leq B$ implies $\< A \>\leq \< B \>$.

Indeed, (E6) follows directly from (E4). For (E7), the assumption gives 
$B-A\geq 0$, hence $\<B-A\>\geq 0$ by (E3), hence 
$\< B \>-\< A \>\geq 0$ by (E2), giving $\< A \>\leq \< B \>$.

An example of a random variable is the number $n$ of eyes on the top 
side of a die. Here $n(\omega)\in\{1,\dots,6\}$ is the number of eyes
on the die visible in experiment $\omega$. We may thus consider $\Ez$
to be the algebra generated by a single random variable $n=n(\omega)$
taking the values $1,2,3,4,5,6$. Thus the relevant random variables are
the functions $A=A(n)$ defined by
\[
A(n)(\omega):=A(n(\omega)).
\]
$A$ is determined by the vector of the six values $A_1=A(1)$,
\ldots, $A_6=A(6)$. Therefore we may identify $\Ez$ with the vector
space $\Rz^{6}$ with componentwise operations. The stochastic model
defined by
\[
\<A\>:=\frac{1}{6}(A_1+\ldots+A_6)
\]
models an ideal, permutation symmetric die.

In practice the set $\Omega$ may be different depending on the
imagination of people and the intended use. The expectation value is
independent of $\omega\in\Omega$ and depends -- as in the example just
given -- on the algebra $\Ez$ of relevant random variables only.
This algebra is always commutative and associative.

A \bfi{statement} is a $\{0,1\}$-valued random variable $A$. The
statement is \bfi{true} (\bfi{false}) in an experiment $\omega$ if
$A(\omega)=1$ (resp. $A(\omega)=0$). The \bfi{probability} of a
statement $A$, defined as
\[
\Pr(A):=\<A\>
\]
is a number between 0 and 1. Indeed, we have $0\le A\le 1$. Thus by
(E3), $\< A \>\geq 0$. By (E6) and (E1), $\< A \>\leq\< 1 \>=1$.

{\bf Proposition.}
If $A_1,\dots,A_n$ are alternative statements of which exactly one is 
true in each experiment then the probabilities $p_{A_{i}} := \Pr(A_i)$ 
sum up to $1$, and $\Pr(A_i\wedge A_j)=0$ for $i\ne j$.

Indeed, the random variable $E:=\D\sum^{n}_{i=1}{A_{i}}$ satisfies
\[
\chi(\omega)=\sum^{n}_{i=1}{A_{i}}(\omega)=1 \Forall \omega\in\Omega,
\]
because by definition exactly one $A_i$ occurs in each experiment
$\omega$. Therefore $E=1$, and
\[
1 = \< 1\>=\<E\>=\Big\<\D\sum^{n}_{i=1}{A_{i}}\Big \>
 = \sum^{n}_{i=1}\<{A_{i}} \>=\sum^{n}_{i=1}p_{A_{i}}.
\]
Similarly, $(A_i\wedge A_j)(\omega)=A_i(\omega)\wedge A_j(\omega)
=0$ for $i\ne j$ since at most one of $A_i(\omega)$ and $A_j(\omega)$
can be true, hence $A_i\wedge A_j=0$ and
$\Pr(A_i\wedge A_j)=\Pr(0)=\<0\>=0$.

The \bfi{cumulative distribution function} (\bfi{CDF}) of a random
variable $A$ is the function $\cdf : \Rz \to [0,1] $ defined by
as
\[
{\cdf}(x):={\Pr}(A \leq x)=\<[A \leq x]\>
\]
for all $x$. Here $[\ldots]$ denotes the statement ($\{0,1\}$-valued
function) defined by the formula inside the square brackets.

{\bf Proposition.}
Every CDF is monotone increasing and satisfies
\lbeq{e.cdf}
\lim_{\eps\downto 0}{\cdf(x+\eps)}=\cdf(x),~~~
\lim_{x \to -\infty}{\cdf(x)}=0,~~~
\lim_{x \to +\infty}{\cdf(x)}=1,~~~
\eeq
Thus every CDF is continuous from the right.

\bepf
By (E6), $x\leq x'$ implies
\[
[A \leq x']-[A\leq x]\ge 0
\]
so by (E6),
\[
\Pr(A\leq x')-\Pr(A \leq x)=\<[A \leq x']-[A\leq x]\>
=\<[x<A \leq x']\>=\Pr([x<A \leq x']) \ge 0.
\]
and thus
\[
{\cdf}( x)\le{\cdf}( x').
\]
To prove continuity from the right we note that the random variable
\[
B(\eps):=[x<A\le x+\eps]
\]
vanishes at $\omega$ with $A(\omega)\le  x$ and for
$\eps<A(\omega)- x$ if $A(\omega)> x$, and is 1 otherwise. Therefore
$B(\eps)\downto 0$ for $\eps\downto 0$. Thus
\[
\<B(\eps)\>\downto 0  \for \eps\downto 0
\]
by (E5). But
\[
\<B(\eps)\>=\Pr(x<A\le x+\eps)=\cdf(x+\eps)-\cdf( x),
\]
hence the first limit of \gzit{e.cdf} follows. The other limits are
proved in the same way.
\epf

Let $A$ be a random variable with CDF $\cdf$.
If $f$ is a step function with finitely many jump points at
$x_1\le\ldots\le x_n$ and $f$ is continuous from the left then
\lbeq{e.st0}
\<f(A)\>= f(x_{1}){\cdf}(x_1) +
\sum_{k=1}^{n-1} f(x_{k+1})({\cdf}(x_{k+1})-{\cdf}(x_k)).
\eeq
Indeed, the right hand side equals the expectation of
\[
f(x_{1})[A\le x_1]+\ldots+\sum f(x_{k})[x_{k-1}<A\le x_k]=f(A).
\]
By taking a continuum limit in \gzit{e.st0}, we obtain the
\bfi{Stieltjes integral representation}
\[
\<f(A)\>=\int^{\infty}_{-\infty} f(\xi) d\cdf(\xi).
\]
for the expectation of $f(A)$, for any function $f$ that is continuous
from the left. Thus the CDF contains all information about expectations
of functions of a real random variable.

For a random variable $A$ with continuously differentiable cumulative
distribution function, the \bfi{distribution} or \bfi{density}
$\rho$ of $A$ is defined as
\[
\rho(\xi)=\frac{d}{d\xi}\cdf(\xi).
\]
It is always nonnegative since the CDF is monotone increasing. The CDF
can be expressed in terms of the density as
\[
{\cdf}(\xi)=\int_{-\infty}^{\xi} \rho(\zeta) d\zeta
\]
If the density $\rho(x)$ exists, it also carries all information about
expectations of functions of $A$. Indeed, we have
\[
\< f(A)\> =\int^{\infty}_{-\infty} f(\xi)\rho(\xi) d\xi
\]
since $d\cdf(\xi)=\rho(\xi)d\xi$.

\subsection{Description dependence of probabilities}\label{ss.condProb}

Classical probabilities are dependent on the description used. 
The latter encode the assumed knowledge about the system under study. 
Note that the implied concept of \bfi{knowledge} is not the knowledge 
of a particular observer or person but an informal shorthand for what 
is modeled in a particular description. Thus the 'knowledge available', 
i.e., the knowledge encoded into a particular description, 
is an objective property of the description used to model the system, 
independent of who 'knows' or 'uses' it. Nothing mental is implied. 
A change of knowledge is therefore just a change of the model used to 
describe a particular system.

Sample expectations and hence sample probabilities have a clear
operational meaning. But they are properties of the specific sample
taken -- changing the sample, e.g., when the values for new 
realizations become known, almost always changes the values of sample
expectations and sample probabilities.

Cumulative distribution functions are easy to estimate on a sample $S$
by the sample CDF
\[
{\cdf}_S(\xi)
=\frac{\mbox{number of $\omega\in S$ with $x(\omega)\le\xi$}}{|S|},
\]
corresponding to the sample expectation \gzit{e.sample}.
A sample CDF is always a step function with discontinuities at the
$x(\omega)$ with $\omega\in S$. In many cases it is well approximated
by a smoothed CDF, using one of many smoothing methods available.
The derivative of the smoothed CDF then serves as an estimate for the
density of a random variable $A$ in a stochastic model in which $A$ is
treated as a random variable with density.

In the univariate case treated above in detail, going from a known 
sample to predictions for not yet known samples is based on the 
possibility to approximate sample distributions of different sample 
sizes by a single model distribution since their CDFs are very similar. 
This also holds in the multivariate case, though different techniques 
must be used to establish corresponding results. 
Finite sample properties can be proved using arguments basically 
similar to our proof of the weak law of large numbers \gzit{e.sigN}; 
under reasoable assumptions, the uncertainty
of the most relevant random variables scales again with $O(N^{-1/2})$.
This is the contents of large $N$ approximations that take the form of
laws of large numbers and central limit theorems. They guarantee
enhanced predictability of the kind that the mean uncertainty is
approximately the single case uncertainty divided by $\sqrt{N}$.

Thus the precise meaning of expectations and probabilities depends on 
which stochastic model is used for a given situation. To see what 
happens when we change the description we consider the concept of 
conditional expectation, which models the reweighing of evidence 
leading to a change of the description of a model. A \bfi{weight} is a 
nonzero random variable $P\ge0$. Let 
\[ 
\Omega_P=\{\omega\in\Omega\mid P(\omega)>0\}
\]
We project the algebra $\Ez$ of random variables 
$X:\Omega\rightarrow\Rz$ to an algebra $\Ez_P$ of random variables 
$Y:\Omega_P\rightarrow\Rz$ by means of the  homomorphism 
\[
\cdot|_{\Omega_P}:\Ez\to\Ez_P: X\mapsto X|_{\Omega_{P}}
\]
that restricts all random variables $X\in\Ez$ to 
$X|_{\Omega_P}\in\Omega_P$. The \bfi{conditional expectation} of 
$X:\Omega\to\Rz$ with respect to the weight $P$ is defined as
\[ 
\< X|_{\Omega_P}\>_P=\< X\>_P:=\frac{\< X\cdot P\>}{\< P\>};
\]
note that $\<P\>>0$ by (E4). The resulting mapping 
$\< \cdot \>_P : \Ez_P\to\Rz$ satisfies all expectation axioms:
(E1)--(E3) and (E5) follow directly from the corresponding axioms for 
$\<\cdot\>$. Only (E4) is nontrivial; $\< X^2 \>_{P}=0$ implies
$\<X^2P\>=\< P\>\< X^2 \>_{P}=0$. But $X^2P\ge 0$, hence (E4) 
gives $X^2P=0$, i.e., $X(\omega)^2P(\omega)=0$ for all $\omega$. Thus 
$X(\omega)=0$ whenever $P(\omega)=1$. This gives $X|_{\Omega_P}=0$
and proves (E4). Therefore $\<\cdot\>_{P}$ is a proper expectation for 
the experiments in $\Omega_{P}$. 

In particular, any statement $A$ with positive probability $\Pr(A)> 0$
may be considered as a weight. Since $\Omega_A$ is the set of all those 
experiments where $A$ is true, conditional expectation with respect to 
$A$ is just expectation in the light of the assumed evidence that
$A$ is true. In particular, the probability for $A$ being true changed
to $1$. More generally, the \bfi{conditional probability} of a statement
$B$ given the statement $A$ is defined by
\[
\Pr({B|} A):=\< B\>_A=\frac{\< BA\>}{\< A\>}
=\frac{\Pr(A\wedge B)}{\Pr(A)};
\]
the last equality holds since
\[ 
BA(\omega)=B(\omega)A(\omega)
=\cases{ 1 & iff $A(\omega)\wedge B(\omega)$, \cr
         0 & otherwise.}
\]
It is easily checked that for any statements $A$ and $B$ we have
\lbeq{3.2}
\Pr(B\wedge A) =\Pr({B|} A)\Pr(A),
\eeq
\lbeq{3.3}
\Pr({A|} B)=\Pr(A)\frac{\Pr({B|} A)}{\Pr(B)}.
\eeq
\gzit{3.3} is often called the \bfi{Bayes theorem}. In this context, 
$\Pr(A)$ is called the \bfi{prior probability} of $A$, $\Pr({A|} B)$ 
its \bfi{posterior probability}, 
and $\Pr({B|} A)/\Pr(B)$ the \bfi{update ratio}. If $A$ and $B$ are 
independent then $\< A B\>=\< A\>\< B\>$, hence 
$\Pr({B|} A)/\Pr(B)=\< A B\>/ \< A\>\< B\>=1$. 
Thus the update ratio captures the degree to which the  knowledge of 
$B$ affects knowledge of $A$. Thus, Bayes theorem allows us to describe 
the change of c-probability of a class of statements when new
information (namely the statement $B$) arrives and is accepted as valid.
Bayes theorem is important to understand what it means to get new 
information when $A$ is a statement of interest and $B$ is 
information that becomes known.  Bayes theorem tells how the 
probability of $A$ changes under  new insight. $\Pr(A)$ changes into 
$\Pr({A|} B)$, so the probability of $A$ must be multiplied by the 
update ratio $\Pr({B|} A)/\Pr(B)$.

\subsection{Deterministic and stochastic aspects}\label{ss.stat}

\nopagebreak
\hfill\parbox[t]{14.6cm}{\footnotesize

{\em Some people are hoping to reintroduce determinism in some way, 
perhaps by means of hidden variables or something like that, but it 
just doesn't work according to the accepted ideas. I might add that
personally I still have this prejudice against indeterminacy in basic 
physics. I have to accept it because we cannot do anything better at 
the present time. It may be that in some future development we shall 
be able to return to determinism, but only at the expense of giving up 
something else, some other prejudice which we hold to very strongly
at the present time.}

\hfill Paul Dirac, 1972 \cite[p.7]{Dir73}
}

\bigskip

The Ehrenfest picture shows that the q-expectations behave like 
classical variables and are governed by a deterministic, Hamiltonian 
dynamics. Thus in the thermal interpretation, quantum physics is a 
deterministic theory. 

Various theorems (beginning with \sca{Bell} \cite{Bel}) and associated 
experiments (beginning with \sca{Aspect} \cite{Asp}) exclude the 
existence of certain kinds of deterministic theories explaining quantum 
mechanics through hidden variables that assume some form of locality, 
noncontextuality, or sharpness; cf. Subsection \ref{ss.nonlocal}.
Since the deterministic Ehrenfest dynamics for q-expectations follows
from the Schr\"odinger equation, there can be no conflict with such 
no-go theorems for deterministic dynamics. Indeed, most q-expectations, 
for example those of momentum or kinetic energy, are nonlocal objects, 
and sharpness is in the thermal interpretation denied from the start. 
The germ of the new interpretation was actually my analysis in 
\cite{Neu.ens} of sharpness arguments in a number of no-go theorems.

A quantity $A$ is considered to be \bfi{significant} if
$\sigma_A\ll|\ol A|$, while it is considered as \bfi{noise} if
$\sigma_A\gg|\ol A|$.
If $A$ is a quantity and $\wt A$ is a good numerical
approximation of its value then $\Delta A:=A-\wt A$ is
noise. Sufficiently significant quantities can be treated as
deterministic; the analysis of noise is the subject of statistics.

Statistics is based on the idea of obtaining information about noisy
quantities of a system by repeated \bfi{sampling} from a
\bfi{population}
of independent systems with identical preparation, but differing in
noisy details not controllable by the preparation. In the present
context, such systems are described by the same Hilbert space, the same
set of quantities to be sampled, and the same state $\<\cdot\>_0$.
The quantities therefore belong to the algebra $\Lin \Hz_S$ of linear
operators on a Euclidean space $\Hz_S$ dense in the Hilbert space of
the system.

More precisely, stochastic features emerge when we consider a large 
sample of similar subsystems of a quantum system. 
For an ensemble of independent measurements on identically prepared
systems, the consensus of all interpretations is that q-expectations
represent (within the accuracy allowed by the law of large numbers)
a statistical average of the measurement results.\footnote{\label{f.erg}
In order to take the q-variance alternatively as a time average of a
single system, one would need to
invoke an ergodic theorem stating that the time average equals the
ensemble average. However such an ergodic theorem makes sense only
semiclassically, and is valid only for very simple classical systems.
Most deterministic systems are far from ergodic.
(This is mentioned, e.g., in the statistical physics book by 
\sca{Landau \& Lifschitz} \cite[Footnote 2, p.12]{LL.5}.)
From the eight examples of statistical models for deterministic systems 
given in Subsection \ref{ss.stochDet} below, only two -- cases (vi) and
(vii) -- have a generally valid ensemble interpretation in terms of 
ergodicity!
Thus the interpretation of the q-variance as a time average is usually
not warranted. This also means ({\sca Neumaier} \cite{Neu.myth}) that
-- in contrast to what is usually done in the popular literature --
the so-called vacuum fluctuations cannot be interpreted as fluctuations
in time.
} 
In order that the thermal interpretation is viable we must therefore be
able to show that the q-variance is indeed a statistical property of
many independent measurements on identically prepared systems.

We regard the systems of the population considered as subsystems of a 
bigger system (e.g., the laboratory) whose set of quantities is given 
by the algebra $\Lin \Hz$ of linear operators on a big Euclidean space 
$\Hz$ dense in the Hilbert space of the big system. To model identically
prepared subsystems we consider injective homomorphisms from $\Lin\Hz_S$
into $\Lin \Hz$ mapping each reference quantity $A\in\Lin\Hz_S$ to the
quantity $A_l \in\Lin\Hz$ of the $l$th subsystem considered to be
`identical' with $A$. Of course, in terms of the big system, the $A_l$
are not really identical; they refer to quantities distinguished by
position and/or time. That the subsystems are \bfi{identically prepared}
is instead modeled by the assumption
\lbeq{e.identical}
\<A_k\> = \<A_l\> \Forall k\ne l,
\eeq
and that they are \bfi{independent} by the assumption
\lbeq{e.independent}
\<A_kA_l\> = \<A_k\>\<A_l\> \Forall k\ne l.
\eeq
The following result is fundamental for statistical considerations:

\bfi{Theorem (Weak law of large numbers).}
For a sample of quantities $A_l$ $(l=1, \ldots , N)$ satisfying
\gzit{e.identical} and \gzit{e.independent}, the \bfi{mean} quantity
\[
  \wh A := \frac{1}{N} \D \sum ^N _{l=1} A_l
\]
(which again is a quantity)
satisfies for any $l$
\lbeq{e.sigN}
\< \wh A \> =\<A_l\>,~~~\sigma_{\wh A} = \sigma_{A_l}/\sqrt{N}.
\eeq

\bepf
By \gzit{e.identical} and \gzit{e.independent},  $\mu:=\<A_l\>$ and
$\sigma:=\sigma_{A_l}$ are independent of $l$, and we have
\[
\<\wh A\> =\frac{1}{N}(\<A_1\>+\dots+\<A_N\> )
=\frac{1}{N}(\mu+\dots+\mu)=\mu,
\]
\lbeq{e.ffsum}
\<\wh A\,^*\wh A\,\>
=\frac{1}{N^2}\Big\<\Big(\sum_jA_j\Big)^*\Big(\sum_kA_k\Big)\Big\>
=N^{-2}\sum_{j,k}\<A_j^*A_k\>.
\eeq
Now
\[
\<A_j^*A_j\>=\<A_j\>^*\<A_j\>+\sigma_{A_j}^2=|\mu|^2+\sigma^2,
\]
and by \gzit{e.independent} for $j\neq k$,
\[
\<A_j^*A_k+A_k^*A_j\>=2\re \<A_j^*A_k\>
=2\re \<A_j\>^*\<A_k\>=2\re \mu^*\mu=2|\mu|^2.
\]
In the sum in \gzit{e.ffsum}, this leads to a contribution of
$|\mu|^2+\sigma^2$ for each of the $N$ diagonal elements, and of
$2|\mu|^2$ for each of the ${N \choose 2}$ pairs of off-diagonal
elements. Therefore
\[
\<\wh A\,^*\wh A\,\>
=N^{-2}\Big(N(|\mu|^2+\sigma^2)+{N \choose 2}2|\mu|^2\Big)
=N^{-1}\sigma^2 + |\mu|^2,
\]
so that
\[
\sigma_{\wh A}^2
= \<\wh A\,^*\wh A\,\>-\<\wh A\,\>^*\<\wh A\,\>
=N^{-1}\sigma^2,
\]
and the assertions follow.
\epf

As a significant body of work in probability theory shows, the
conditions under which $\sigma_{\wh A}\to 0$ as $N\to\infty$ can
be significantly relaxed; thus in practice, it is sufficient if
\gzit{e.identical} and \gzit{e.independent} are approximately valid.

The significance of the weak law of large numbers lies in the fact that
\gzit{e.sigN} becomes arbitrarily small as $N$ becomes sufficiently
large. Thus the uncertainty of quantities when averaged over a large
population of identically prepared systems becomes arbitrarily small
while the mean value reproduces the value of each quantity.

The weak law of large numbers implies that, in a context where
many repeated experiments are feasible, states can be given a
\bfi{frequentist} interpretation, in which $\ol A=\<A\>$ is the
\bfi{expectation} of $A$, empirically defined as an average over
many realizations. In this case (and only in this case), the uncertainty
$\sigma_A$ becomes the standard deviation of $A$; then it
captures the absolute accuracy of the individual realizations.

Note that there are also many single systems with a sound statistical
interpretation. Once one has a time series of a single system that
empirically looks fluctuating, one can do valid statistics with it.
Sun spot activity data or El Nino data (both time series for single
physical systems) are traditional test data for statistical procedures.
Single deterministic systems data with highly oscillating contributions
-- like the weather on Earth -- are routinely explained and forecast in
terms of a stochastic process. Thus, whenever the statistics of 
interest is that of a system's mean behavior in time (or in space), 
it is meaningful to do statistics on the single system. The time 
(or space) coordinates then label the sample under consideration.
(See Footnote ${}^{\ref{f.erg}}$ for the related notion of ergodicity.)

On the other hand, in equilibrium thermodynamics, where a tiny number
of macroscopic observations on a single system completely determine
its state to engineering accuracy, such a frequentist interpretation
is inappropriate.

Thus, the thermal interpretation captures correctly the experimental
practice, and determines the conditions under which
deterministic and statistical reasoning are justified:

\bfi{(SP)} \bfi{Statistical principle:}
{\em Deterministic reasoning is appropriate for all sufficiently
significant quantities.
Statistical reasoning is necessary for noisy quantities, and requires
that these quantities are sufficiently similar and sufficiently 
independent to ensure that their mean is significant.} 

In this way, statistical considerations naturally arise from the
nonstatistical foundations, giving \bfi{statistics} the same role as in
classical physics, namely as the {\em art and science of interpreting
measurements}. Real experiments are (and must be) designed such
that they allow one to determine approximately the relevant properties
of the state under study, hence the values of all quantities of
interest.
The uncertainties in the experiments imply approximations, which,
if treated probabilistically, need an {\em additional} probabilistic
layer.
Expectations from this secondary layer, which involve probabilistic
statements about situations that are uncertain due to neglected
but in principle observable details (cf. \sca{Peres} \cite{Per}),
happen to have the
same formal properties as the values on the primary layer, though
their physical origin and meaning is completely different.

\subsection{What is probability?}\label{ss.whatIsProb}

In Subsection \ref{ss.probViaEx}, we derived from (E1)--(E5) the 
traditional probabilistic machinery for single real random variables. 
More generally, it can be proved (see \sca{Whittle} \cite{Whi}) that 
even in the multivariate case,
the above approach to classical probability is equivalent to the measure
theoretical approach in the traditional axiomatic setting of Kolmogorov.
In this equivalence, $\Omega$ is an abstract measure space, a
stochastic model is a probability measure on $\Omega$, and $\Ez$ is a
vector space of random variables with finite expectation.


The exposition in \sca{Whittle} \cite{Whi} (or, in more abstract
terms, already in \sca{Gelfand \& Naimark} \cite{GelN}) shows that,
if the $X_j$ are pairwise commuting, it is possible to define
for any Gibbs state in the present sense, random variables $X_j$
in Kolmogorov's sense such that the expectation of all sufficiently
regular functions $f(X)$ defined on the joint spectrum of $X$
agrees with the value of $f$. It follows that in the pairwise commuting
case, it is always possible to construct a probability interpretation
for the quantities, completely independent of any assumed microscopic
reality. (If the components of $X$ do not commute, a probabilistic
interpretation in the Kolmogorov sense is no longer possible because of
the nonclassical uncertainty relations \gzit{e.uncRel}.)

The details (which the reader unfamiliar with measure theory may
simply skip) are as follows. We may associate with every vector $X$
of quantities with commuting components a time-dependent, monotone
linear functional $\<\cdot\>_t$ defining the \bfi{expectation}
\[
      \<f(X)\>_t:=\Tr \rho(t) f(X)
\]
at time $t$ of arbitrary bounded continuous functions $f$ of $X$.
These functions define a commutative $C^*$-algebra $\Ez(X)$.
The \bfi{spectrum} $\spec X$ of $X$ is the set of all $*$-homomorphisms
(called \bfi{characters}) from $\Ez(X)$ to $\Cz$, and has
the structure of a Hausdorff space, with the \bfi{weak-$*$ topology}
obtained by calling a subset $S$ of $\spec X$ closed if, for any
pointwise convergent sequence (or net) contained in $S$, its limit is
also in $S$. Now an expectation functional satisfying (E1)--(E5) turns 
out to be equivalent to a multivariate probability measure $d\mu_t(X)$
(on the sigma algebra of Borel subsets of the spectrum $\Omega$ of $X$)
defined by
\[
   \int d\mu_t(X) f(X) := \sint \rho(t) f(X) =\<f(X)\>_t.
\]
Both Whittle's and Kolmogorov's foundation of classical probability 
theory are axiomatic and hence independent of the 
interpretation of the axioms. We may refer to probability as defined
by Kolmogorov or Whittle as \bfi{c-probability}. 

In generalization of this, we refer, for any Hermitian operators $P$ 
satisfying $0\le P\le 1$, to its expectation as the \bfi{q-probability}
\[
\Pr(P):=\<P\>
\]
of $P$. As a special case, we call a quantity $P$ satisfying 
$P^2=P=P^*$ a \bfi{statement}; then $0\le P\le 1$ follows by the 
spectreal theorem. For a statement $P$, the uncertainty of its 
probability $p=\ol P$ is $\sigma_P=\sqrt{p(1-p)}$ since by 
\gzit{e.sigmaA}, $\sigma_P^2=\<P^2\>-\ol P^2=p-p^2$.
Another special case is the q-probability of a self-adjoint Hermitian 
q-observable $A$ taking values in some open interval $]a,b[$ of real
numbers, defined as
\[
\Pr(A\in{]a,b[}):=\<P_{]a,b[}(A)\>,
\]
where $P=P_{]a,b[}(A)$ is the spectral projector of $A$ to the interval 
$]a,b[$. Note that here $P^2=P=P^*$, so that $P$ is a statement --
the formal equivalent of the informal statement '$A$ is in $]a,b[$'.

Whittle's approach is essentially equivalent to the commutative case of 
the formal core of quantum mechanics, interpreted in statistical terms.
Then q-probabilities and c-probabili\-ties agree.

We discuss in more detail the important special case of binary tests, 
where Born's rule frequently applies essentially exactly.
An \bfi{ideal binary measurement}, e.g., the click of a detector, is 
described by a statement $P$ coding the presence (1) or absence (0) of 
a click. In particular, a \bfi{test for a state}\footnote{
Note that a test for $\phi$ turns out positively with probability 1 if 
the measured system is in the pure state $\phi$. However, it also turns
out positively with a positive probability if the measured system is
in a pure state different from $\phi$, as long as it is not
orthogonal to it. Thus calling it a 'test for $\phi$', though
conventional, is something of a misnomer.
} 
$\phi$ with $\phi^*\phi = 1$ is an ideal binary measurement of 
$P=\phi\phi^*$; it is easily checked that this is a statement.
By the above, such a test turns out positive with probability
$p = \<P\>$. In particular, if the system is in a
pure state $\psi$ then $p =\<P\>= \psi^*P\psi=\psi^*\phi\phi^*\psi
= |\phi^*\psi|^2$, hence 
\[
p = |\phi^*\psi|^2.
\]
This is the well-known \bfi{squared probability amplitude} formula
appearing in the scattering form of Born's rule as stated in Part I 
\cite[Subsection 3.1]{Neu.Ifound}. Thus the scattering form of Born's 
rule appears as natural consequence rather than as a basic axiom.

\subsection{Probability measurements}\label{ss.probMeas}

\nopagebreak
\hfill\parbox[t]{10.8cm}{\footnotesize

{\em When it uses probabilities, [...] science regards them [...] as 
measurable (and calculable) physical quantities like lengths, energies, 
and wavelengths. [...] The probability of a truly single event is 
intrinsically unmeasurable and [...] science has nothing to say about''}
it. {\em ''To obtain the value of a physical quantity, one must measure 
it a number of times, Each measurement contains an error, and the 'true'
 value is (usually) computed as the arithmetic mean of all measured 
values. [...] in a similar way''}, we {\em ''measure the relative 
frequency [...] of an event in a series of trials. Each relative 
frequency contains an error, and the 'true' probability is computed as 
the mean of the relative frequencies over a number of series. [...] 
nothing strange or inconsistent is left in the idea of probability as a 
measurable physical quantity.
}

\hfill Henry Margenau, 1950 \cite[pp.250--252]{Mar}
}

\bigskip

By its definition, the notions of q-expectations and q-probabilities 
belong to the formal core of quantum mechanics and are independent of 
any interpretation. But in the thermal interpretation all 
q-expectations, and in particular all q-probabilities, are among the 
beables. We discuss here basic aspects of their measurement;
a more thorough discussion of measurement from the point of view of the 
thermal interpretation is given in Part III \cite{Neu.IIIfound}.

By the law of large numbers, q-expectations can be measured with in
principle arbitrarily high accuracy by taking sample means of low 
accuracy measurements, whenever there is a device (the 
\bfi{preparation}) -- e.g., a particle accelerator -- that produces a 
large number of independent copies (realizations) of the same quantum 
system. The accuracy improves by a factor of $\sqrt{N}$, where $N$ is 
the sample size. 

In the same way, the q-probability $p$ are approximately measurable as 
relative frequencies. 
As a consequence of the weak law of large numbers \gzit{e.sigN}, the 
uncertainty of the \bfi{relative frequency} $p_N$, defined as the 
sample mean of ideal binary measurements in a sample of $N$ independent 
realizations of the statement $P$, is 
$\sigma=\sigma_P\sqrt{N}=/\sqrt{p(1-p)/N}$. This uncertainty 
approaches zero when the sample size $N$ gets arbitrarily large.  
Thus measuring a probability by a relative frequency gives in principle 
arbitrarily accurate results. 

This gives a fully adequate operational definition of probabilities
without any logical problems, of the same quality as operational
definitions of highly accurate length or time measurements:
The q-probabilities are theoretical observables; they are measured as 
relative frequencies, to some reasonable accuracy that can be 
quantified by the associated uncertainty. We draw conclusions about 
sufficiently uncertain situations based on observed sample means and 
relative frequencies on a sample of significant size, and we quantify 
our remaining uncertainty by statistical safeguards (confidence 
intervals, etc.), well knowing that these sometimes fail.
For example, the \bfi{$5\sigma$-rule} for the discovery of elementary 
particles tries to guard against such failures. 
This view of probabilities as measurable entities is just the one 
described in the above quote by Margenau (if 'final' is read for 
'true').

Thus probability has an objective interpretation precisely to the 
extent that objective protocols for taking the sample measurements
i.e., how to distinguish a positive from a negative test, are agreed 
upon. Any subjectivity remaining lies in the question of deciding which
protocol should be used for accepting a measurement as 'correct'.
Different protocols may give different results.
Both classically and quantum mechanically, the experimental context
needed to define the protocol influences the outcome.
In particular, there is a big difference between the description of an 
event before ({\em predicting}) 
or after ({\em analyzing}) it occurs. This is captured rigorously in 
classical probability theory by conditional probabilities (cf. 
Subsection \ref{ss.condProb}), and less rigorously in quantum physics by
the so-called collapse of the wave function, where the description of 
the state of a particle is different before and after it passes a 
filter (polarizer, magnet, double slit, etc.). Thus we may view the 
\bfi{collapse} as the quantum analogue of the change of conditional 
probability when the context changes due to new information.
Recognizing this removes another piece of strangeness from quantum 
physics.

\subsection{The stochastic description of a deterministic system}
\label{ss.stochDet}

A stochastic description of a deterministic system is a reduced
deterministic description by moments rather than details. Formally,
it is obtained by restricting an algebra of commuting, classical 
quantities describing a given deterministic system to the subalgebra of 
quantities completely symmetric in some properties declared 
\bfi{indistinguishable}
for the purposes of the reduced (or coarse-grained) description. In the
simplest case, where expectation is defined as sample expectation, the
individual realizations over which the sample mean is taken are declared
indistinguishable, with the consequence that only symmetric functions
of relatizations -- hence functions of expectations -- are available
in the reduced descriptions.

In the terminology of knowledge discussed in Subsection 
\ref{ss.condProb}, such a reduction amounts to forgetting or 
ignoring information known from the more detailed model. We have 
additional modeling uncertainty due to the lack of detail in the 
description used. This description is independent of a probabilistic 
interpretation.\footnote{
though via the construction of Subsection \ref{ss.whatIsProb} below, 
it can always be given one in terms of c-probabilities
} 
It just means that one only considers a limited family of well-behaved 
relevant quantities in place of the multitude of quantities in a more 
detailed description.

The analysis presentented here allows one to apply statistical models to
complicated {\em deterministic} situations -- not only in physics -- 
and to {\em single} complicated spatial events or time series.
In each case, a suitable concept of expectation is introduced that,
as a figure of speech, allows one to make probabilistic and other
statistical statements about deterministic situations.

Important examples of  statistical models for deterministic situations
with increasingly random appearance are: 

(i)
The deterministic but irregular sequence of prime numbers
(\sca{Tennenbaum} \cite{Ten}). 
Here experiments are the natural numbers, and expectations are 
introduced through a mathematically rigorous limit.

(ii)
Rounding errors in deterministic floating-point computations
(\sca{Vignes} \cite{Vig}). 
Here experiments are sequences of floating-point operations, and 
expectations are introduced through an empirical model for
single rounding errors and an assumption of independence.

(iii)
Texture in a single (hence fully determined) picture
(\sca{Heeger \& Bergen} \cite{HeeB}). 
Here experiments are hypothetical images of the same size, and 
expectations are introduced through a mean over a (not well-defined) 
neighborhood of pixels.

(iv)
Economic time series, e.g., prices of oil and electricity
(\sca{Granger \& Newbold} \cite{GraN}). 
Here experiments are hypothetical scenarios for the time series, and 
expectations are introduced through a no longer fully well-defined
time average.

(v)
Tomorrow's weather prediction, when based on classical fluid mechanics
(\sca{Gneiting \& Raftery} \cite{GneR}). 
Here experiments are again hypothetical weather scenarios, and 
expectations are introduced through an even less well-defined 
space-time average.

(vi)
Deciding for red or black by spinning a roulette wheel
(\sca{Hopf} \cite{Hopf}). 
Here expectations may be introduced through symmetry arguments (for an
ideal wheel), or through ergodic theory.

(vii)
The classical statistical mechanics of an ideal gas. 
Here \sca{Boltzmann} \cite{Bol} introduced expectations through an 
average over all particles.

(viii)
The classical statistical mechanics of solids and fluids. 
Here \sca{Gibbs} \cite{Gib} introduced expectations through a 
fictitious average over many systems with identical macroscopic 
properties.

\section{The thermal interpretation of quantum field theory}
\label{s.qft}

In the thermal interpretation, the fundamental description of reality 
is taken to be standard relativistic quantum field theory.\footnote{
It does not matter whether or not there is a deeper underlying structure
such as that of string theory, in terms of which quantum field theory
would be an effective theory only.
} 
On the formal (uninterpreted) level, the formal core of quantum physics
is valid for both quantum mechanics and quantum field theory. But since
the algebra of quantities considered is different, there are two 
essential differences between quantum mechanics and quantum field 
theory:

First, in place of position and momentum of finitely many particles in
quantum mechanics one has in quantum field theory operators for fields.
Each field $\phi$ has a space-time dependence that satisfy Galilei or
Poincare invariance and causal commutation relations. Thus each field
provides an infinitude of uncertain quantitites. More precisely, since
from a rigorous point of view, field operators $\phi(x,t)$ at spatial
position $x$ and time $t$ are distribution-valued operators, the
quantities in quantum field theory are \bfi{smeared fields}, local
space-time integrals 
\[
\phi(f)=\int_\Omega f(x,t)^T\phi(x,t)dxdt
\]
over local patches $\Omega$ in space-time, where $f$ is a smooth 
\bfi{test function} (e.g., a Gaussian), and multipoint generalizations 
of these.

Second, unlike in quantum mechanics, position in quantum field theory
is not an operator but a parameter, hence has no associated uncertainty.
The uncertainty is instead in the quantities described by the details 
of the test functions $f$ associated with real field measurements.

In this section we show that these differences strongly affect the 
relation between quantum field theory and reality. Among the beables
of quantum field theory (discussed in Subsection \ref{ss.beablesQFT})
are smeared field expectations and pair correlation functions, which 
encode most of what is of experimental relevance in quantum field 
theory. Subsection \ref{ss.dynQFT} comments on relativistic quantum 
field theory at finite times, a usually much neglected topic essential 
for a realistic interpretation of the universe in terms of quantum 
field theory, given in Subsection \ref{ss.universe}.
Finally we discuss notions of causality (Subsection \ref{ss.causal}) 
and nonlocality (Subsection \ref{ss.nonlocal}) and their relation to 
the thermal interpretation.

\subsection{Beables and observability in quantum field theory}
\label{ss.beablesQFT}

The most directly observable (and hence obviously real) features of 
a macroscopic system modeled by quantum field theory are the 
q-expectations of smeared versions of the most important quantum fields,
integrated over cells of macroscopic or mesoscopic size.

This identification can be made because statistical mechanics
allows one to derive for the q-expectations of the fields the equations
of state of equilibrium thermodynamics for cells of macroscopic size in
thermal equilibrium and the hydromechanical equations for cells of
mesoscopic size in local equilibrium. Both are known to yield excellent
macroscopic descriptions of matter.

For macroscopic systems, one must necessarily use a coarse-grained
description in terms of a limited number of parameters. In the quantum
field theory of macroscopic objects, any averaging necessary for 
applying the law of large numbers is already done inside the definition 
of the macroscopic (i.e., smeared) operator to be measured. As shown in 
statistical mechanics, this is sufficient to guarantee very small 
uncertainties of macroscopic q-observables.
Thus one does not need an additional averaging in terms
of multiple experiments on similarly prepared copies of the system.
This is the deeper reason why quantum field theory can make accurate
predictions for single measurements on macroscopic systems.

In addition to the macroscopic ontology just described, the thermal 
interpretation also has a microscopic ontology concerning the reality 
of inferred entities. As in our earlier discussion of quantum mechanics,
the thermal interpretation declares as real but not directly
observable any q-expectation $\<A(x,t)\>$ of operators and any
\bfi{q-correlation}, the q-expectation of a product of operators at 
pairwise distinct points. More precisely, the q-expectations 
$\<\phi(x,t)\>$ of fields are distributions that produce the -- 
in principle approximately measurable -- numbers $\<\phi(f)\>$ when 
integrated over sufficiently smooth localized test functions $f$. 
Certain q-correlations are also measurable in principle by probing the 
state with external fields in linear or nonlinear response 
theory.
See, e.g., \sca{H\"anggi \& Thomas} \cite{HaenT} for classical 
correlations and \sca{Davies \& Spohn} \cite{DavS} for quantum 
correlations.

Scattering experiments provide further observable information, about 
time-ordered multi-point correlations of these fields. The related 
S-matrices also appear in microlocal kinetic descriptions of dilute 
macroscopic matter at the level of the Boltzmann equation or the 
Kadanoff--Baym equations. These are derived from the q-expectations 
of products of fields at two different space-time points.
(The kinetics of the Boltzmann equation derived from the particle
picture has long been replaced by more accurate Kadanoff--Baym
equations derived from field theory.)

2-point correlations in quantum field theory are effectively 
classical observables; indeed, in kinetic theory they appear as the 
classical variables of the Kadanoff-Baym equations, approximate 
dynamical equations for the 2-point functions. After a Wigner transform
and some further approximation (averaging over small cells in phase 
space), these turn into the classical variables of the Boltzmann 
equation. After integration over momenta and some further approximation 
(averaging over small cells in position space), these turn into the 
classical variables of the Navier-Stokes equation, hydromechanic 
equations that describe the behavior of macroscopic fluids. 
For macroscopic solids, one can use similar approximations to arrive 
at the equations of elasticity theory. The most detailed classical 
level, the Kadanoff-Baym equations, still contain the unsmeared 
ensemble means of field products.

According to the thermal interpretation, there is nothing in quantum 
field theory apart from q-expectations of the fields and q-correlations.
The quantities accessible to an observer are those q-expectations and
q-correlations whose arguments are restricted to the observer's 
world tube. More precisely, what we can observe is contained in the 
least oscillating contributions to these q-expectations and 
q-correlations.
The spatial and temporal high frequency part is unobservable due to
the limited resolution of our instruments.

All macroscopic objects are objects describable by hydromechanics 
and elasticity theory; so their classical variables have the same 
interpretation. Thus the quantum-mechanical ensemble averages are 
classical variables. Moreover, because of the law of large numbers, 
$\<f(x)\> \approx f(\<x\>)$ for any sufficiently smooth function 
$f$ of not too many variables. (These caveats are needed because high 
dimensions and highly nonlinear functions do not behave so well under 
the law of large numbers.) In particular, we get from Ehrenfest's 
theorem \gzit{e.Ehrenfest} the standard classical Hamiltonian 
equations of motion for macroscopic objects.

Statistical mechanics shows that the uncertainties in the
macroscopically relevant smeared fields scale with the inverse square
root of the cell volume. This means that integrals over $\<\Phi(x,t)\>$ 
are macroscopically meaningful only to an accuracy of order $V^{-1/2}$ 
where $V$ is the volume occupied by the mesoscopic cell containing $x$, 
assumed to be homogeneous and in local equilibrium. 
This is the standard assumption for deriving from first
principles hydrodynamical equations and the like. It means that the
interpretation of a field gets more fuzzy as one reduces the amount of
coarse graining -- until at some point the local equilibrium
hypothesis is no longer valid.

Everything deduced in quantum field theory about macroscopic
properties of matter in local equilibrium or dilute matter in the
kinetic regime follows, and one has a completely self-consistent
setting. The transition to classicality is automatic and needs no deep
investigations: The classical situation is simply the limit of a
huge number of particles, where the law of large numbers discussed in
Subsection \ref{ss.stat} reduces the uncertainty to a level below
measurement accuracy.

\bigskip

Since quantum fields are quantities depending on a space-time argument,
one can prepare or measure events at any particular space-time position 
at most once. Thus it is impossible to repeat measurements, and the 
standard statistical interpretation in terms of sufficiently many 
identically prepared systems is impossible. Therefore, the notion of 
ensemble cannot be understood as an actual repetition by repeated 
preparation.

In particular, the measurement of quantum fields is not covered by 
Born's rule in its standard measurement-based form. This is the reason 
why (unlike scattering applications) macroscopic applications of 
quantum field theory never invoke Born's rule. 
Indeed, we had concluded in  Part I \cite{Neu.Ifound} that at finite
times (i.e., outside its use in interpreting asymptotic S-matrix
elements), Born's rule cannot be strictly true in relativistic quantum
field theory, and hence not in nature.

\subsection{Dynamics in quantum field theory}\label{ss.dynQFT}

The description of dynamics in current relativistic quantum field 
theory textbooks
is a delicate subject. In many such books, physical meaning is given 
only to scattering processes, i.e., the behavior at asymptotic times 
$t\to\pm\infty$, whose statistical properties are expressed in terms 
of time-ordered correlation functions. Textbooks commonly restrict 
their attention to the calculation of the low order contributions to 
the scattering amplitudes and how these are renormalized to give finite 
results. Questions about the quantum field dynamics at finite times are 
not discussed since the dynamics is deeply buried under the formal 
difficulties of the renormalization process needed to make relativistic 
quantum field theory work. Sometimes they are even claimed to be 
impossible to address!\footnote{
For example, \sca{Scharf} \cite[introduction to Chapter 2]{Scharf.QED} 
writes: 
''The more one thinks about this situation, the more one is led to the 
conclusion that one should not insist on a detailed description of the 
system in time. From the physical point of view, this is not so 
surprising, because in contrast to non-relativistic quantum mechanics, 
the time behavior of a relativistic system with creation and 
annihilation of particles is unobservable. Essentially only scattering 
experiments are possible, therefore we retreat to scattering theory. 
One learns modesty in field theory.''
} 
However, on the level of rigor customary in theoretical physics,
quantum field dynamics at finite time is actually well-defined in terms 
of the so-called closed time path (CTP) approach; see, e.g., 
\sca{Calzetta \& Hu} \cite{CalH.book}.

Since the traditional Schr\"odinger picture breaks manifest Poincar\'e
invariance, relativistic QFT is almost always treated in the Heisenberg
picture. The Heisenberg dynamics on the fields is given by
\[
\frac{\partial}{\partial x_\nu} \phi(x)=p_\nu\lp \phi(x),
\]
where $p_\nu$ is the $\nu$th operator component of the \bfi{4-momentum}
vector $p$, defined as the generator of the translations of the 
Poincar\'e group. In particular, the physical Hamiltonian $H=cp_0$, 
where $c$ is the speed of light, is obtained after the construction of 
the $N$-point functions (q-expectations of fields and q-correlations) 
as the operator generating the time shift of the fields.

In the context of the thermal interpretation, a natural covariant
generalization of the Ehrenfest equation \gzit{e.Ehrenfest} is obtained
by considering, in place of the time-dependent q-expectations in the 
nonrelativistic setting discussed before,
q-expectations dependent on a space-time location $x$ in
Minkowkski space, and to require that the resulting space-time dependent
q-expectations $\<A\>_x$ satisfy the \bfi{covariant Ehrenfest equation}
\lbeq{e.covEhrenfest}
\frac{\partial}{\partial x_\nu} \<A\>_x=\<p_\nu\lp A\>_x.
\eeq
One easily concludes that for arbitrary space-time points 
$x,y,z,w,\ldots$,
\[
\<\phi(z)\>_x=\<\phi(z+x-y)\>_y,
\]
generalizing the nonrelativistic \gzit{e.timeShift}, and 
\[
\<\phi(z)\phi(w)\cdots \>_x=\<\phi(z+x-y)\phi(w+x-y)\cdots\>_y.
\]
Thus the complete spacetime-dependence of q-expectations, and in 
particular their dynamics, is determined by the q-expectations at any 
particular fixed space-time position.

From the covariant Ehrenfest picture we may now deduce a
\bfi{covariant Schr\"odinger picture}, by writing
\[
\<A\>_x:=\Tr \rho(x)A
\]
with a space-time dependent density operator $\rho$. Then
\gzit{e.covEhrenfest} becomes the \bfi{covariant von Neumann equation}
\[
i\hbar\frac{\partial}{\partial x_\nu} \rho(x)=[p_\nu,\rho(x)].
\]
The rank is preserved. Hence if $\rho$ has rank 1, this is equivalent
with writing $\rho(x)=\psi(x)\psi(x)^*$, where $\psi(x)$ satisfies
the \bfi{covariant Schr\"odinge equation}
\[
i\hbar\frac{\partial}{\partial x_\nu} \psi(x)=p_\nu\psi(x).
\]
After rescaling, we see that in quantum field theory, time 
$t_\nu=x_\nu/c$ and energy $H_\nu=cp_\nu$ -- not space and momentum -- 
have become 4-dimensional:
\[
i\hbar\frac{\partial}{\partial t_\nu} \psi(x)=H_\nu\psi(x).
\]

\subsection{The universe as a quantum system}\label{ss.universe}

Our solar system can be approximately treated classically, but from the
fundamental point of view of quantum field theory it must be considered
as a quantum system. The state of the solar system, when modeled by
quantum fields, completely specifies what happens in any small
space-time region within the solar system.

Traditional interpretations of Copenhagen flavor require that a quantum
system is measured by an external classical apparatus. They cannot
apply to the quantum field theory of our solar system, say, since we 
do not have access to an external classical apparatus for measuring 
this system. The astronomers doing measurements on the solar system
are part of the system measured -- a situation outside the Copenhagen
setting. The thermal interpretation has no such problems.

The thermal interpretation is even consistent with assigning a 
well-defined (though only superficially known) state to the whole 
universe, whose properties account for everything observable within 
this universe.

Unlike in conventional single-world interpretations of quantum
physics, nothing in the thermal interpretation depends on the
existence of measurement devices (which were not available in the very
far past of the universe). Thus the thermal interpretation allows one
to consider the single \bfi{universe} we live in as a quantum system, 
the smallest closed physical system containing us, hence strictly 
speaking the only system to which unitary quantum physics applies 
rigorously.

There is no longer a physical reason to question the existence of the
state of the whole universe, even though all its details may be unknown
for ever. Measuring all q-observables or finding its exact state
is already out of the question for a small macroscopic quantum system
such as a piece of metal. Thus, as for a metal, one must be content
with describing the state of the universe approximately.

What matters for a successful physics of the universe is only that
we can model (and then predict) those q-observables of the universe that
are accessible to measurement at the temporal and spatial scales of
human beings. Since all quantities of interest in a study of the
universe as a whole are macroscopic, they have a tiny uncertainty and
are well-determined even by an approximate state. For example, one could
compute from a proposed model of the universe the uncertain values
of the electromagnetic field at points where we can measure it, and
(if the computations could be done) should get excellent agreement
with the measurements.

Since every q-observable of a subsystem is also a q-observable of the
whole system, the state of the universe must be compatible with
everything we have ever empirically observed in the universe.
This implies that the state of the
universe is highly constrained since knowing this state amounts to
having represented all physics accessible to us by the study of its
subsystems. This constitutes a very stringent test of adequacy of a
putative state of the universe.

Cosmology studies the state of the universe in a very coarse (and partly
conjectured) approximation where even details at the level of galaxies
are averaged over. Only for q-observables localized in the solar system
we have a much more detailed knowledge.

Of course, a more detailed discussion of the state of the universe
should include gravitation and hence would touch on the difficult,
unsolved problem of quantum gravity. However, the thermal interpretation
gives at least a consistent interpretational framework in which to
discuss these questions - without having to worry about whether the
concepts used to formulate these questions mean anything in the context
of a quantum system without external observers.

In relativistic quantum field theory, the basic fields are local
objects in the sense that smeared fields $\phi(f)$ and $\phi(g)$
commute whenever $f$ and $g$ have spacelike separated support.
Nevertheless, a quantum state specifies the q-expectations $\<\phi(f)\>$
for arbitrary smooth test functions $f$ and the higher order moments
($n$-point correlation functions) $\<\phi(f_1)\ldots \<\phi(f_n)\>$
for arbitrary smooth test functions $f_1,\ldots,f_n$, hence makes
statements about uncertain values at all space-time locations.
Hence relativistic quantum field theory necessarily describes a
complete universe. When gravitation is not modeled, everything happens 
in a Minkowski spacetime.

\subsection{Relativistic causality}\label{ss.causal}

\nopagebreak
\hfill\parbox[t]{10.8cm}{\footnotesize

{\em Phrases often found in the physical literature as 'disturbance
of phenomena by observation' or 'creation of physical attributes of
objects by measurements' represent a use of words like 'phenomena' and
'observation' as well as 'attribute' and 'measurement' which is hardly
compatible with common usage and practical definition and, therefore,
is apt to cause confusion. As a more appropriate way of expression,
one may strongly advocate limitation of the use of the word phenomenon
to refer exclusively to observations obtained under specified
circumstances, including an account of the whole experiment.
}

\hfill Niels Bohr, 1948 \cite{Bohr}
}

\bigskip

\nopagebreak
\hfill\parbox[t]{10.8cm}{\footnotesize

{\em In each experiment, irrespective of its history, there is only
one quantum system, which may consist of several particles or other
subsystems, created or annihilated at the various interventions.
}

\hfill Asher Peres and Daniel Terno, 2002 \cite[p.98]{PerT1}:
}

\bigskip

We now consider relativistic causality in a Minkowski spacetime. By
working with charts, everything said generalizes to the case of curved
spacetime. Quantized spacetime is not discussed since there is no
accepted framework for it.

A \bfi{point object} has, at any given time in any observer's frame,
properties only at a single point, namely the point in the intersection
of its world line and the spacelike hyperplane orthogonal to the
observer's 4-momentum at the time (in the observer frame) under
discussion.
An \bfi{extended object} has properties that depend on more than one
spacelike-separated space-time position.
A \bfi{joint property} is a property that explicitly depends on more
than one space-time location within the space-time region swept out by
the extended object in the course of time.

Note that from a fundamental point of view there is no clear demarcation
line that would tell when a system of particles (e.g., a molecule, or
a sollar system) should or should not be regarded as a single object.
The thermal interpretation therefore treats arbitrary subsystems of a
large system as a single object if they behave in some respect like a
unity.

We may distinguish three Poincar\'e invariant definitions of causality.

\pt
\bfi{Point causality:} Properties of a point object depend only on its
closed past cones, and can influence only its closed future cones.
This is used in special relativity, which discusses the motion of a
single classical particle in a classical external field.

\pt
\bfi{Separable causality:} Joint properties of an extended object
consist of the combination of properties of their constituent points.
This is intuitively assumed in all discussions of Bell-type nonlocality,
and is in conflict with experiments involving highly entangled photons.

\pt
\bfi{Extended causality:} Joint properties of an extended object depend
only on the union of the closed past cones of their constituent parts,
and can influence only the union of the closed future cones of their
constituent parts. This is the version that can probably be derived
from relativistic quantum field theory, where particles are localized
excitations of the quantum field, and hence extended objects.

All three notions of causality agree on the causality properties of
point objects ('point causality') but differ on the causality
properties of extended objects. 
If one regards an entangled quantum system as a system of point
particles one runs into lots of counterintuitive conceptual problems.
If one regards an entangled quantum system as a single extended system
in the above sense, all such difficulties disappear. 

Extended causality is the form of causality appropriate for the thermal 
interpretation. It takes into account what was known almost from the 
outset of modern quantum physics - that quantum objects are 
intrinsically extended and must be treated as whole. 

The extended system view gives the appropriate intuition.
The violation of Bell inequalities in experiments such as those by
\sca{Aspect} \cite{Asp} (cf. Subsection \ref{ss.nonlocal}) show 
that neither point causality or separable causality can be realized in 
nature. But extended causality is not ruled out by current experiments. 
\sca{Eberhard \& Ross} \cite{EbeR} gives a proof of causality from 
relativistic quantum field theory, in the sense that no faster than 
light communication is possible.

\subsection{Nonlocal correlations and conditional information}
\label{ss.nonlocal}

Bell's theorem, together with  experiments that prove that Bell 
inequalities are violated imply that reality modeled by deterministic
process variables is intrinsically nonlocal. The thermal interpretation
explicitly acknowledges that all quantum objects (systems and 
subsystems) have an uncertain, not sharply definable (and sometimes 
extremely extended) position, hence are intrinsically nonlocal. 
Thus it violates the assumptions of Bell's theorem and its variations.

Here we briefly discuss some related aspects of nonlocality from the 
point of view of the thermal interpretation.

We use the example of an entangled 2-photon state. In quantum physics, 
there is a definite concept of a system in a 2-photon state,
but only a fuzzy one of 'two photons'. Attempting to literally
interpret the two photons in a system with an entangled 2-photon state 
leads to paradoxes related to seemingly acausal nonlocal correlations.

Given the quantum mechanical 2-photon system together with the 
Schr\"odinger dynamics determined by the associated dispersion relation,
Born's rule makes assertions about measurements anywhere in
the universe at any future time! Something more nonlocal cannot be
conceived. It is therefore no surprise that intuition is violated. 
However, the thermal interpretation renounces the  universal validity 
of Born's rule. Instead, the nonlocal correlations get their natural 
explanation in terms of extended causality and conditional information. 

Consider the conceptual setting of a typical Bell-type experiment, 
where entangled 2-photon systems are subjected to measurements by two
far away observers conventionally called Alice and Bob, at times and
positions corresponding to spacelike distances. Their measurement
results are later compared by another observer called Charles, say.

In an experiment checking Bell inequalities, an
object in the form of a 2-photon system may be prepared at the source by
parametric down-conversion and propagating freely in two opposite
directions. Whatever Alice and Bob measure far away depends on the
whole 2-photon system. According to the thermal interpretation, the
object described by this 2-photon system is an extended object.
Over long distances, the uncertainty intrinsic to the
2-photon system becomes huge. The object becomes vastly extended -- so
nonlocal that the assumptions in Bell's argument are obviously violated.
It is not surprising that the conclusions can be violated, too.

Any meaningful use of the notion of causality depends on a notion of
before and after, which does not exist in the case of correlations at
spacelike distances. 
Because of the Lorentz invariance of all relativistic arguments,
one cannot say that Alice's actions and observations cause (or affect)
Bob's observations to be correlated once  -- as usually assumed --
Alice's and Bob's position are causally unrelated. For in this case,
there are always Lorentz frames in which Alice acts later than Bob
observes, and others in which Bob acts later than Alice. So neither can 
be said to cause (or affect) the other observer
(cf. \sca{Peres \& Terno} \cite{PerT1,PerT2}).
Whatever statistics can be made (by Bob or Charles) from data collected 
by Bob before Alice's choices or results become available to Charles --
it will be completely unaffected by the behavior of Alice and her 
detector.

But something else from Alice becomes known to Bob faster than light --
conditional information. \bfi{Conditional information} is information 
deduced from what -- given past and present observations available to 
a local observer --  is known from theory but
is not observed itself by this observer. Additional observations may
make conditional information more precise. But as long as part of the
data in the condition is not yet known, nothing conclusive is known.
Having about tomorrow's weather the conditional information that 
'Should there be no clouds it will not rain' tells in fact nothing 
useful about the weather tomorrow, unless we have information about 
tomorrow's clouds.

Similarly, Bob gets conditional information of the kind: 'Should Alice
have measured X then her result was Y.' Because Bob does not know
whether the hypothesis holds, he knows nothing useful. 
Bob's claimed knowledge about the results of Alice's measurement is
sound only if Alice actually measured something. If she instead took a
nap, or if her detector failed because of a power outage, Bob concluded
something wrongly.

Causality only demands that information flow is limited by the speed
of light. Nothing in
relativity forbids conditional information to be passed faster than
light. For example, we know lots of conditional information about what
can or cannot happen inside black holes although no information can
flow out from there. Such conditional information is obtained from
theory independent of observation. But theoretical conclusions apply
instantaneously and have no speed limit.

In Bell-type experiments, the conditional information and the
correlations become actual only when someone (like Charles) has access
to the actual data resolving the condition. It is easily seen that 
extended causality is observed. 

One may ask, however, how it is 
possible that, in actual long-distance entanglement experiments 
performed in the past, when Charles checked the findings he found 
Alice's conditonal information unconditionally satisfied. The reason 
for this is that, when preparing actual long-distance entanglement 
experiments, the experimenters make sure that nothing will interrupt the
expected flow of events needed for a correct experimental performance. 
This means that they in fact prepare not only the source of the
entangled photon pairs but also Alice's and Bob's environment and the 
whole environment the photons traverse during the experiment. This
preparation must be careful enough to exclude all events that would 
cause unexpected changes to the intended protocol or remove the 
delicate entanglement. But this means that the preparation deposits
in the past light cones of both Alice and Bob enough correlated 
classical information that influences the present of Alice and Bob 
when they perform their choices. This doesn"t explain everything about
the observed correlations but casts some doubt on the validity of the 
stringent assumptions made in derivations of Bell-type inequalites.

\section{Conclusion}

The thermal interpretation 

\pt
acknowledges that there is only one world;

\pt
has no split between classical and quantum mechanics -- the former
emerges naturally as the macroscopic limit of the latter;

\pt
is by design compatible with the classical ontology of ordinary
thermodynamics;

\pt
is description-dependent but observer-independent, hence free from
subjective elements;

\pt
is about both real systems and idealized systems, at every level of
idealization;

\pt
applies both to single quantum objects (like a quantum dot, a neutron 
star, or the universe) and to statistical populations;

\pt
satisfies the principles of locality and Poincare invariance, as defined
in relativistic quantum field theory;

\pt
is compatible with relativistic extended causality;

\pt
uses no concepts beyond what is taught in every quantum physics course;

\pt
involves no philosophically problematic steps.

No other interpretation combines these merits. The thermal
interpretation leads to a gain in clarity of thought. This results in
saving a lot of time otherwise spent in the contemplation of aspects
arising in traditional interpretations, which from the point of view of
the thermal interpretation appear as meaningless or irrelevant side
problems.

The thermal interpretation of quantum physics (including quantum
mechanics, statistical mechanics and quantum field theory) is an
interpretation of everything. It allows a consistent quantum
description of the universe from the smallest to the largest levels of
modeling, including its classical aspects.

These foundations are easily stated and motivated since they are
essentially the foundations used everywhere for uncertainty
quantification, just slightly extended to accommodate quantum effects
by not requiring that q-observables commute.

\bigskip

In all traditional interpretations of quantum physics, the theoretical
'observables' are unobservable operators. It is no surprise that
calling unobservable things observables causes apparent indeterminism.
It was a historical misnomer that lead to the strange, unfortunate
situation in the foundations of quantum physics that persists now for
nearly hundred years.

On the other hand, in the thermal interpretation of quantum physics,
the theoretical observables (the beables in the sense of Bell) are the 
expectations and functions of them. They satisfy a deterministic 
dynamics. Some of these beables are practically (approximately) 
observable. In particular, q-expectations of Hermitian quantities and
q-probabilities, the probabilities associated with appropriate 
self-adjoint Hermitian quantities, are among the theoretical 
observables. The q-expectations are approximately measured by
reproducible single measurements of macroscopic quantities, or by
sample means in a large number of observations on independent, similarly
prepared microscopic systems. The q-probabilities are approximately
measured by determining the relative frequencies of corresponding 
events associated with a large number of independent, similarly 
prepared systems. 

This eliminates all foundational problems that were introduced into 
quantum physics by basing the foundations on an unrealistic concept of
observables and measurement. With the thermal interpretation, the
measurement problem turns from a philosophical riddle into a
scientific problem in the domain of quantum statistical mechanics,
namely how the quantum dynamics correlates macroscopic readings from
an instrument with properties of the state of a measured microscopic
system. This problem will be discussed in Part III \cite{Neu.IIIfound}.

\bigskip

\end{document}